\documentclass[sigconf]{acmart}

\AtBeginDocument{%
  \providecommand\BibTeX{{%
    \normalfont B\kern-0.5em{\scshape i\kern-0.25em b}\kern-0.8em\TeX}}}

\acmBooktitle{}

\usepackage{booktabs}  
\usepackage[utf8]{inputenc}

\usepackage{microtype}

\usepackage{tabu}
\usepackage[table]{xcolor}
\usepackage{colortbl}

\usepackage[dvipsnames]{xcolor}
\definecolor{thedarkblue}{RGB}{0,0,120} %
\definecolor{mydarkblue}{rgb}{0,0.08,0.45} %
\definecolor{darkblue}{rgb}{0,0.08,180}
\colorlet{TufteRed}{red!80!black}
\definecolor{theblue}{RGB}{0,0,180}
\colorlet{thered}{TufteRed}
      
\usepackage{microtype}
\usepackage{balance}

\usepackage{booktabs}
\usepackage{tabularx}

\newcommand{\eat}[1]{\ignorespaces}
\usepackage{comment}

\usepackage{tikz}
\usepackage{verbatim}
\usetikzlibrary{arrows}
\usetikzlibrary{shapes,snakes}
\usetikzlibrary{decorations.pathmorphing} %
\usetikzlibrary{fit}					%
\usetikzlibrary{backgrounds}	%

\usepackage{ragged2e}
\usepackage{makecell}
\usepackage{multirow}
\usepackage{microtype}
\usepackage{balance}
\usepackage{setspace}

\graphicspath{{./}{./graphics/}}
\newcolumntype{H}{>{\setbox0=\hbox\bgroup}c<{\egroup}@{}}

\newcolumntype{R}[1]{>{\RaggedLeft\arraybackslash}} %
\newcolumntype{L}[1]{>{\RaggedRight\arraybackslash}} %

\AtBeginEnvironment{pmatrix}{\setlength{\arraycolsep}{2pt}}

\DeclareMathOperator{\hugeE}{\mbox{\huge\raise-0.3ex\hbox{E}}}
\DeclareMathOperator{\p}{\mathbb{P}}
\DeclareMathOperator{\hugep}{\mbox{\huge\raise-0.3ex\hbox{$\p$}}}

\definecolor{googleblue}{HTML}{4285F4}
\definecolor{googlered}{HTML}{DB4437}
\definecolor{googlepurple}{HTML}{A142F4} %
\definecolor{googlegreen}{HTML}{0F9D58}

\usepackage{float}
\usepackage{graphbox}
\usepackage{svg}
\usepackage{nicefrac}

\usepackage{bold-extra}
\usepackage[T1]{fontenc}

\usepackage{array}
\newcolumntype{P}[1]{>{\centering\arraybackslash}p{#1}}
\newcolumntype{M}[1]{>{\centering\arraybackslash}m{#1}}

\definecolor{orange}{rgb}{1,0.5,0}
\definecolor{graynode}{RGB}{20,20,20}
\definecolor{crimsonred}{RGB}{220,20,60}
\definecolor{darkgraynode}{gray}{0.5}
\definecolor{lightgraynode}{gray}{0.8}

\usepackage{rotate}
\usepackage{adjustbox}
\usepackage{array}
\usepackage{capt-of}
\usepackage{tabulary}
\usepackage{setspace}
\usepackage{mathtools}
\usepackage{pifont} 
\usepackage{multirow}
\usepackage{makecell}   
\usepackage[capitalize,noabbrev]{cleveref}

\definecolor{gray}{RGB}{20,20,20}
\definecolor{gray}{RGB}{0.7,0.7,0.7}
\definecolor{greencm}{RGB}{0,153,0}

\definecolor{plotblue}{RGB}	{30,144,255}
\definecolor{plotgreen}{RGB}	{50,205,50}
\definecolor{plotred}{RGB}	{220,20,60}

\definecolor{myyellow}{RGB}{255,255,204}
\definecolor{myred}{RGB}{255,204,204}
\definecolor{myblue}{RGB}{0,200,255}
\definecolor{mygreen}{RGB}{80,220,80}

\newcommand*\hrulefillvar[1][0.4pt]{\leavevmode\leaders\hrule height#1\hfill\kern0pt}

\usepackage{nicefrac}

\DeclareMathAlphabet{\mathbcal}{OMS}{cmsy}{b}{n}
\usepackage{mathrsfs}
\usepackage{comment}

\usepackage{bm}
\usepackage{bbm}

\usepackage{paralist}
\definecolor{thedarkblue}{RGB}{0,0,120} 
\definecolor{mydarkblue}{rgb}{0,0.08,0.45} 
\definecolor{salmon}{RGB}{232,125,114}

\usepackage{hyperref}
\hypersetup{%
    colorlinks=true,
    linkcolor=mydarkblue,
    citecolor=mydarkblue,
    filecolor=mydarkblue,
    urlcolor=mydarkblue}

\usepackage{subfigure}

\usepackage{dblfloatfix}
\usepackage{tcolorbox}
\usepackage{fontawesome5}

\newtcolorbox{promptbox}[2][]{
  colback=white!98!blue!2, 
  colframe=blue!70!black, 
  coltitle=white,          
  fonttitle=\bfseries\sffamily,
  title={\faTasks[regular]\hspace{1mm}~#2},
  left=2mm, right=2mm, top=1mm, bottom=1mm,
  arc=3mm,                  
  boxrule=0.9pt,
}

\begin{document}
\emergencystretch 1.5em

\title{MLLM as a UI Judge: Benchmarking Multimodal LLMs for Predicting Human Perception of User Interfaces}

\author{Reuben A Luera}
\affiliation{%
  \institution{University of California, Berkeley}
  % \city{Berkeley}
  % \state{CA}
  % \country{USA}
  \country{}
}

\author{Ryan Rossi}
\affiliation{%
  \institution{Adobe Research}
  % \city{San Jose}
  % \state{CA}
  % \country{USA}
  \country{}
}

\author{Franck Dernoncourt}
\affiliation{%
  \institution{Adobe Research}
  % \city{Seattle}
  % \state{WA}
  % \country{USA}
  \country{}
}

\author{Samyadeep Basu}
\affiliation{%
  \institution{Adobe Research}
  % \city{San Jose}
  % \state{CA}
  % \country{USA}
  \country{}
}

\author{Sungchul Kim}
\affiliation{%
  \institution{Adobe Research}
  % \city{San Jose}
  % \state{CA}
  % \country{USA}
  \country{}
}

\author{Subhojyoti Mukherjee}
\affiliation{%
  \institution{Adobe Research}
  % \city{San Jose}
  % \state{CA}
  % \country{USA}
  \country{}
}

\author{Puneet Mathur}
\affiliation{%
  \institution{Adobe Research}
  % \city{San Jose}
  % \state{CA}
  % \country{USA}
  \country{}
}

\author{Ruiyi Zhang}
\affiliation{%
  \institution{Adobe Research}
  % \city{San Jose}
  % \state{CA}
  % \country{USA}
  \country{}
}

\author{Jihyung Kil}
\affiliation{%
  \institution{Adobe Research}
  % \city{San Jose}
  % \state{CA}
  % \country{USA}
  \country{}
}

\author{Nedim Lipka}
\affiliation{%
  \institution{Adobe Research}
  % \city{San Jose}
  % \state{CA}
  % \country{USA}
  \country{}
}

\author{Seunghyun Yoon}
\affiliation{%
  \institution{Adobe Research}
  % \city{San Jose}
  % \state{CA}
  % \country{USA}
  \country{}
}

\author{Jiuxiang Gu}
\affiliation{%
  \institution{Adobe Research}
  % \city{San Jose}
  % \state{CA}
  % \country{USA}
  \country{}
}

\author{Zichao Wang}
\affiliation{%
  \institution{Adobe Research}
  % \city{San Jose}
  % \state{CA}
  % \country{USA}
  \country{}
}

\author{Cindy Xiong Bearfield}
\affiliation{%
  \institution{Georgia Institute of Technology}
  % \city{Atlanta}
  % \state{GA}
  % \country{USA}
  \country{}
}

\author{Branislav Kveton}
\affiliation{%
  \institution{Adobe Research}
  % \city{San Jose}
  % \state{CA}
  % \country{USA}
  \country{}
}
    
\renewcommand{\shortauthors}{Reuben Luera, et al.}

\begin{abstract}
In an ideal design pipeline, user interface (UI) design is intertwined with user research to validate decisions, yet studies are often resource-constrained during early exploration. Recent advances in multimodal large language models (MLLMs) offer a promising opportunity to act as early evaluators, helping designers narrow options before formal testing. Unlike prior work that emphasizes user behavior in narrow domains such as e-commerce with metrics like clicks or conversions, we focus on subjective user evaluations across varied interfaces. We investigate whether MLLMs can mimic human preferences when evaluating individual UIs and comparing them. Using data from a crowdsourcing platform, we benchmark GPT-4o, Claude, and Llama across 30 interfaces and examine alignment with human judgments on multiple UI factors. Our results show that MLLMs approximate human preferences on some dimensions but diverge on others, underscoring both their potential and limitations in supplementing early UX research.
\end{abstract}


\maketitle

\section{Introduction}

Design choices are inseparable from usability and experience. 
Nowhere is this more evident than in interface design, which shapes user perception of product usability and trustworthiness~\cite{reinecke2013predicting}.
It can even profoundly impact user productivity, and their emotional responses to products~\cite{bhandari2017effects, bailey1988effects}. 
For this reason, user experience research to evaluate interface design is crucial for development~\citep{steen2007early, baxter2015understanding}. 
Through research-informed iterations upon design choices, products can be refined to be more usable, engaging, and trustworthy.

However, many organizations face a persistent bottleneck, where too few researchers are available to keep pace with the growing scale of design and evaluation needs. 
In fact, according to a report from the Nielsen Norman Group, there is only one researcher for every five designers~\cite{kaplan2020}. 
As a result, many organizations are forced to allocate their limited research capacity selectively, testing only a small subset of ideas rather than engaging in rapid, iterative evaluation, which has long been shown to significantly enhance design quality~\cite{medlock2018rapid}.

Existing work in this space simulates human behavior in domains like e-commerce and solely focuses on user metrics like click-through and conversion rates \citep{wang2025agenta}, which can be limited because such metrics capture only general user outcomes and fail to provide diagnostic insights into design quality, usability, or user experience. 
To address this gap, artificial intelligence offers a promising avenue for building more efficient and scalable evaluation pipelines.
Several existing works have examined how large language models (LLMs) can be used to evaluate UIs, such as generating feedback based on critiques and quality ratings from professional designers ~\citep{duan2024uicrit, duan2024generating}. 
While LLMs can be trained to generate feedback on UIs, prior work shows that they often struggle to faithfully capture human responses to visual interfaces~\cite{wang2024aligned, stokes2025write}. 
We build on this work by systematically comparing LLM evaluations with human-generated UI evaluation data, in order to identify both reliable alignments and critical misalignments that highlight future research opportunities. Through this comparison, we identify the LLM prompting approach that most effectively approximates human judgments.

Further, the analysis presented in this paper, specifically how state-of-the-art models perform moderately well in UI judging tasks, can lay the groundwork for using multimodal large language models (MLLMs) to supplement early-stage user research. The results produced can be summarized in the following way: 1) Task 1, the Likert-scale task produced human and MLLM evaluation scores that, when compared to one another, show promise for MLLMs as UI judge approximators.  The models we use (Claude 3.5 Sonnet, GPT-4o, and Llama-3.2-11B-Vision-Instruct) all have an accuracy $\pm1$ score greater than 75\%. Further, we evaluated performance using Pearson, Spearman, and Kendall tau correlation statistics that signify a moderately strong correlation between MLLM and human scores. 2) Task 2, the pairwise comparison task, reflected that MLLMs, when given questions that have large human score differences, can accurately simulate human data. In these pairwise experiments, we saw largely that models' performance improved as the degree of difference between two screens increased. Given this, MLLMs are not at a stage where they should be seen as replacements for real-life human evaluation, but instead can be used to supplement early UI evaluations, especially in scenarios where UIs would remain untested.

\noindent\textbf{Summary of Main Contributions.} 
The key contributions of this work are as follows:

\begin{compactitem}
    \item \textbf{Novel MLLM as a judge evaluation tasks:} We introduce a new task that can be used to evaluate whether large language multimodal models can approximate human preferences and provide reasonable justifications of those preferences in Likert scale evaluations of UIs and pairwise comparisons of real-world interfaces—moving beyond behavioral simulation to subjective UI alignment.
    
    \item \textbf{Absolute Human/MLLM UI judgments benchmark\footnote{\label{benchdata}See supplementary materials for benchmark data}:} Task 1 (\cref{sec:exp-score-pred}) produced 
    a benchmark dataset of user interface evaluations, derived from both MLLM and human evaluators. This dataset was created by showing humans and three state-of-the-art MLLMs (Claude 3.5 Sonnet, GPT-4o, and Llama-3.2-11B-Vision-Instruct) the same set of UIs and then asking them to use a list of cognitive, perceptual, and emotional factors to evaluate them on a scale from 1-7.
    
    \item  \textbf{Pairwise Human/MLLM UI judgments benchmark\footref{benchdata}:} Task 2 (\cref{sec:exp-pairwise-comparison}) produced a benchmark of pairwise MLLM and human evaluations of user interfaces. To create this data, we showed the same three state-of-the-art MLLMs two UIs at a time and prompted them to choose their preferred screen based on the same list of factors. The MLLM data was then compared to human pairwise data (generation techniques shown in \cref{sec:exp-pairwise-comparison}) to calculate another dataset of agreement scores. 

    \item \textbf{MLLM training data:} The data produced in this study can be used as the benchmark to create a fine-tuned model. Training MLLMs and LLMs with novel datasets, such as the one created in this study, has proven to increase human accuracy in MLLM as a judge tasks \citep{kim2024prometheus, ziegler2020finetuninglanguagemodelshuman, lee2024prometheus}.

    \item \textbf{MLLM as a UI judge use cases:} Data analysis that suggests that MLLMs are a capable tool to supplement user interface testing for both absolute score testing and pairwise comparison to gauge perception at early stages. Our data suggest that teams with limited resources can use MLLMs to supplement, rather than fully replace, human UI testing. 
\end{compactitem}

The novelty of this research lies in the exploration of whether or not multimodal LLMs can evaluate \textit{visual} user interfaces, not just text or other behavioral logs. Past LLM as a judge work has largely centered on comparing human and LLM textual evaluations \citep{das2025leveraging}; we instead focus our work on visual UI evaluation. Through this study, we evaluate whether or not MLLMs are capable of emulating human UI judgments and report when they do so well and when they fail. In doing so, we explore to what degree MLLMs can be used as early-stage, low-cost approximators that act as a supplement, not as a replacement, to existing user research.

\section{Related Work}\label{sec:related-work}

\subsection{LLM-as-a-Judge}
Extensive work has been done in the field of 'LLM as a Judge', and within this paper, we aim to build upon the existing literature. LLM as a judge is a concept that can be applied to a myriad of different domains \citep{gu2024survey} and consists of using LLMs to evaluate the outcomes of complex tasks. They can do so effectively because they can quickly and efficiently process extensive datasets that can then be used to make an informed decision \citep{achiam2023gpt}. LLMs have been used as judges to evaluate everything from other models \citep{li2023generative, wang2023large,zhang2023wider, zheng2023judging}, natural language and information \citep{wang2022self, ramamurthy2022reinforcement, ouyang2022training}, charts \citep{kim2025chart}, and reasoning and thinking \citep{achiam2023gpt, team2023gemini, guo2025deepseek}. Despite this, related literature suggests that using LLMs as judges can offer cost-cutting and a flexible alternative to human evaluation, but warns that they often lack consistency while suffering from issues pertaining to bias and reliability \citep{gu2024survey}. 

Across all studied domains, larger state-of-the art models (e.g., GPT-4o, Claude, etc.) perform better in LLM as a judge tasks than smaller models \citep{thakur2025judgingjudgesevaluatingalignment, zheng2023judging}. Nonetheless, even well-performing models deviate by up to five points on a 10-point scale, and LLM evaluators are often excessively verbose, biased towards the existing positions, and prefer their own outputs \citep{liu2024aligning, panickssery2024llmevaluatorsrecognizefavor}. Given these limitations, LLMs often are capable as approximators, but are far from being human replacements in evaluation tasks. 

We aim to distinguish our work by using a method that specifically explores how MLLMs can be used to evaluate two distinct UIs. In the context of this paper, we use a pairwise MLLM evaluation method \citep{liu2024aligning}, which consists of having the MLLM compare two different options and selecting which is better aligned with a predetermined criterion \citep{gu2024survey, qin2024largelanguagemodelseffective}. This method closely aligns with A/B testing, commonly used in user research to compare two interfaces. We aim to run a pairwise test with human users looking at UIs and a Likert-scale LLM evaluation with the same interfaces to explore how close LLMs are to simulating human results.

    \begin{figure*}[h]
    \centering
    \includegraphics[width=.9\linewidth]{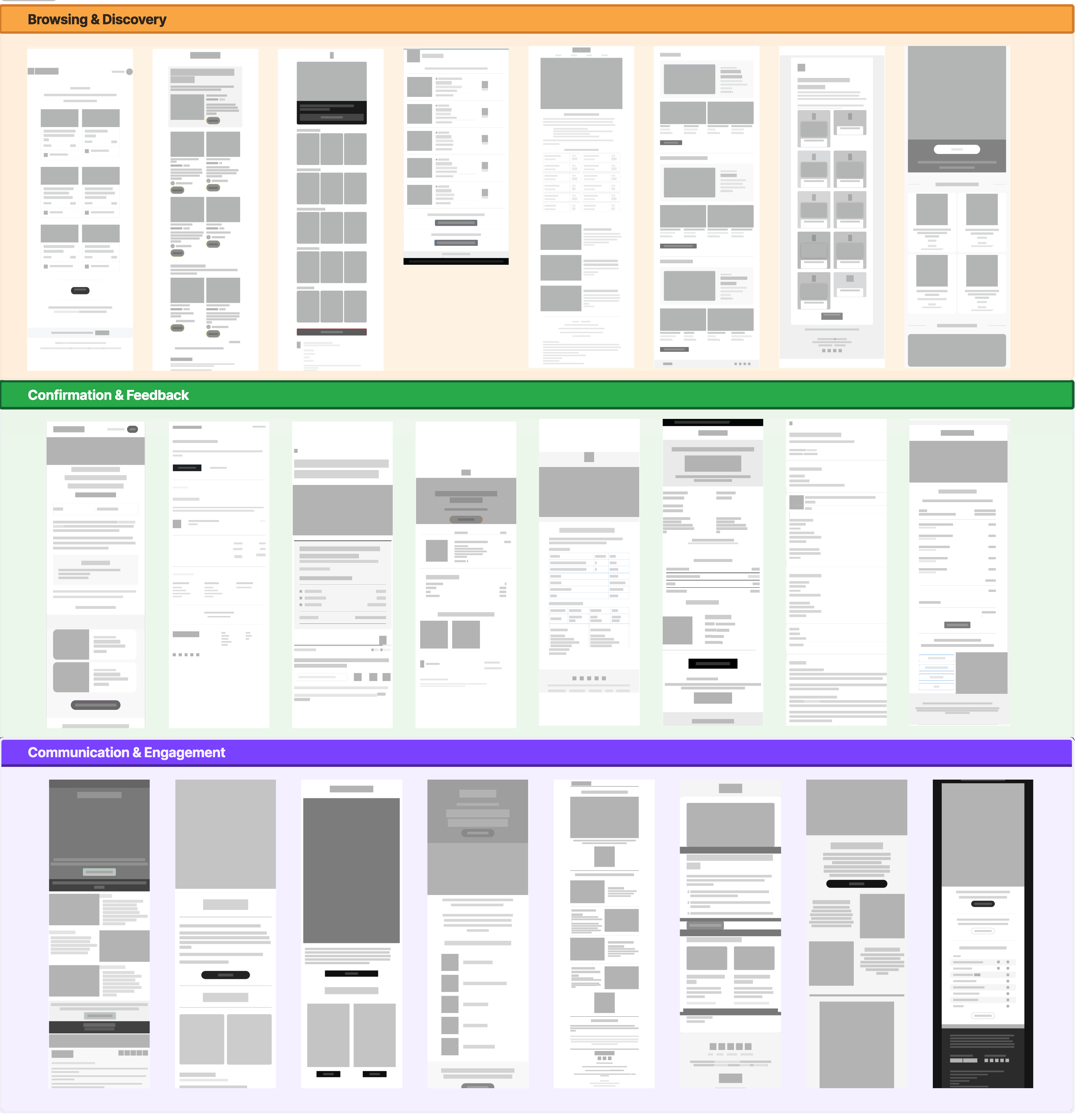}
    \caption{%
    {\textbf{User interfaces split by domains:} 
    An overview of a sample of the UIs 
    evaluated by humans and MLLMs can be divided into three domains: landing pages, digital receipts, and catalogs. Here we present low-fidelity versions of the screens that we presented to the users. The users saw high-fidelity, branded, and in-product versions of these screens that have been anonymized.}}
    \label{fig:Domains}
    \end{figure*}

\subsection{MLLMs in User Research and Design}
As in many other domains, AI is becoming increasingly intertwined with user interface design and research \citep{bertao2021artificial, luera2024survey}. Designers and researchers alike are using MLLM to create and explore different interfaces for their respective products.
Specifically in design research, LLMs are being used as assistants that are able to collaborate during UX evaluation sessions \citep{10.1145/3613904.3642168}. LLMs are occasionally being used to create synthetic data that emulates human users \citep{cao2025survey, rosala2024synthetic, 10.1145/3701716.3715452}. For instance, \citet{DBLP:journals/nlpj/JansenJS23} explain how LLMs can use simulated respondents to supplement the creation and analysis of survey research. Further, \citet{10.1145/3544548.3580688} demonstrates that LLMs are capable of simulating user survey responses and creating responses that humans found plausible. Similarly, \citet{li2024frontiers} found that in larger-scale market research surveys, LLM-generated survey responses achieved about a 75\%-85\% agreement with results from actual humans. However, this paper also highlights the importance of these models being trained on human data, as they cannot yet capture all of the nuances of human users.

There have been some work focused on some aspect of UI evaluations~\citep{duan2024generating, duan2024uicrit, wu2024uiclip,Schoop_2022}. In \citet{duan2024generating}, a Figma plugin was created by assessing GPT-4's alignment with a set of design heuristics. Similarly, \citet{duan2024uicrit} utilized a dataset of design critiques from several experienced designers to improve LLM-generated UI feedback by 55\%. Meanwhile \citet{wu2024uiclip} creates a design tool based on several datasets and shows it performs closely to human ground truth. In addition, \citet{10.1145/3613904.3642481} created an LLM tool used to evaluate smartwatch interfaces using simulated users.
Our work differs from these works as we create a benchmark that compares several different state-of-the-art MLLMs to each other and to the human judgments. Further, the benchmark reflects how MLLMs perform in both pairwise and absolute scoring tasks based on a list of nine UI factors.
Agent A/B \citep{wang2025agenta} asks a similar question and explores whether or not its own model can complete A/B UX tests. They train their agent specifically on UX click-through rate, focusing solely on the filters and menus of e-commerce websites. We aim to focus on a broader question, exploring UI judgments at a higher level while testing an assortment of widely used MLLMs (e.g., GPT-4o, Claude, Gemini) on an assortment of UI screens from different domains.

\section{Study Methodology} \label{sec:ui-to-exp}

\subsection{UI Collection} \label{sec:ui-to-exp-collection}
In order to quickly curate a diverse set of user interfaces, we used pre-made interfaces from a professional UI dataset\footnote{\url{https://reallygoodemails.com/}}. We selectively identified 30 UIs (24 professionally made and 6 made from scratch) across three different categories: browsing \& discovery, confirmation \& feedback, and communication \& engagement (\cref{fig:Domains}) to cover a diverse set of 
UI components, aesthetic styles, content complexity, quality, and purposes. 
For instance, communication \& engagement UIs typically have large images and eye-catching text in order to grab the attention of the user. Meanwhile, screens from the confirmation \& feedback category typically have more text because their purpose is to quickly provide payment information, and catalogs often have pairs of images and text. We aimed to choose UIs that could work on both mobile and desktop devices and put constraints on the height so that most screens were about similar dimensions. For the most part, all UIs were extracted as is from the dataset. Since these screens were made by professional designers, we reduced the build quality of some screens to maintain a diverse dataset of both high and low quality designs.
Since the UIs were collected from a professionally made dataset, we anticipated that most scores, both human and AI, would skew higher and more positively. For this reason, we designed our pairwise and ranking tasks with stronger granularity to detect signals in our measurements.

\subsection{Factors} \label{sec:ui-to-exp-measures}
\begin{table}[t]
    \centering
    \small
    \setlength{\tabcolsep}{0.3em}
    \renewcommand{\arraystretch}{1.0}
    \caption{UI factors and corresponding statements. }
    \begin{tabu}{l ll}
    \toprule
   \textbf{Category} 
   & \textbf{Factor}      & \textbf{Question} \\ 
    \midrule
   
    & \textcolor{googlegreen}{Ease of Use}  & The UI looks \textbf{easy to use.} \\
   \textcolor{googlegreen}{\textsc{\bfseries \scshape Cognitive}} 
    & \textcolor{googlegreen}{Clarity}   & The layout is \textbf{uncluttered}. \\
    & \textcolor{googlegreen}{Visual Hierarchy}  & The UI has a clear \textbf{visual hierarchy}. \\
   \midrule
   
    & \textcolor{googleblue}{Memorable}  &  The UI is easily \textbf{remembered}. \\
    \textcolor{googleblue}{{\bfseries \scshape Perceptual}} 
    & \textcolor{googleblue}{Trust}  &  The UI appears \textbf{trustworthy}.\\
    & \textcolor{googleblue}{Intuitive}  & The UI is \textbf{intuitive}. \\
    \midrule
   
    & \textcolor{googlered}{Aesthetic Pleasure}  & The UI is \textbf{aesthetically} pleasing. \\
   \textcolor{googlered}{{\bfseries \scshape Emotional}} & \textcolor{googlered}{Interest}   & The UI is \textbf{interesting}. \\
    & \textcolor{googlered}{Comfort}  & I feel \textbf{comfortable} with the UI. \\
    \bottomrule
    \end{tabu}
    \label{table:Factors}
\end{table}

The UIs we gathered will be judged on nine UI factors on a Likert Scale from 1-7 \citep{kim2025chart}. These factors were decided upon after consulting literature that explains how people react to designs and visualizations \citep{seeley2012hearing, 8809393}. The nine factors are clustered into three themes: Cognitive, Perceptual, and Emotional, following the design framework described by Seeley et al.~\cite{seeley2012hearing}. 

\subsubsection{\textit{\textbf{\textcolor{googlegreen}{\textsc{\bfseries \scshape Cognitive Factors: }}}}}

\begin{itemize}
    \item \textit{\textbf{\textcolor{googlegreen}{\textsc{\bfseries \scshape Ease of use }}}}refers to how easy an interface appears to be. Ease of use can depend on factors like UI element placement, layout of a design, and design heuristics like the match between the system and the real world \citep{kumar2004user, nielsen2005ten}. A screen with high ease of use often does not make users think; its function is straightforward and obvious.
    
    \item \textit{\textbf{\textcolor{googlegreen}{\textsc{\bfseries \scshape Clarity }}}}pertains how clear and uncluttered an interface looks. For example, if an interface has too many UI elements, it can overwhelm the user by increasing their cognitive load \citep{oviatt2006human, darejeh2024critical}. For this reason, \cite{nielsen2005ten} highlights the importance of ensuring that designs remain minimalist in nature with enough white space. Doing so ensures the creation of a screen that is uncluttered and does not overwhelm the users.
    
    \textit{\textbf{\textcolor{googlegreen}{\textsc{\bfseries \scshape Visual Hierarchy }}}}consists of ensuring that there is a clear order of importance between all of the UI elements on a screen \citep{wang2024research, still2018web}. Doing so decreases the user's cognitive load, as their attention is caught quickly by whatever is largest, most central, and most important. An interface with good visual hierarchy will have elements that are sized and placed based on their perceived importance and relevance to the user.
\end{itemize}

\subsubsection{\textit{\textbf{\textcolor{googleblue}{\textsc{\bfseries \scshape Perceptual Factors: }}}}}\label{sec:perceptual-factors}
\begin{itemize}
    \item \textit{\textbf{\textcolor{googleblue}{\textsc{\bfseries \scshape Memorable }}}}pertains to how easily a UI can be remembered and speaks to the impact that a design has on a user. Memorability can be both negative and positive, and can be a result of attributes like color or recognizable elements \citep{6634103}. Further, does the UI have a certain uniqueness that makes it stand out for the user. These elements, in total, contribute to making an interface memorable to the user. 

    \item \textit{\textbf{\textcolor{googleblue}{\textsc{\bfseries \scshape Trust }}}}pertains to the amount of confidence that the user interface elicits within the user. Headlines, text content, as well as even how well an interface is put together, strongly contribute to trust in interface design. A trustworthy interface will make users feel confident in the content, often because they are designed intentionally and carefully \citep{zieglmeier2021designing}.

    \item \textit{\textbf{\textcolor{googleblue}{\textsc{\bfseries \scshape Intuitive }}}} refers to how functional an interface looks to a user. Intuitiveness and functionality often are intertwined with user experience, so intuitiveness is a crucial factor when judging interfaces. An intuitive screen would be defined as one that users can look at and immediately understand the natural purpose and function \citep{baerentsen2000intuitive, blackler2005intuitive, macaranas2015intuitive}
\end{itemize}

\subsubsection{\textbf{\textcolor{googlered}{\textsc{\bfseries \scshape Emotional Factors:}}}}

\begin{itemize}
    \item \textit{\textbf{\textcolor{googlered}{\textsc{\bfseries \scshape Aesthetic Pleasure }}}}refers to how visually attractive an interface is to a specific user. It is a largely subjective factor that often differs from user to user. For the most part, aesthetic pleasure is tightly tied by choices pertaining to color, font, content density, and overall structure. An aesthetically pleasing interface is able to integrate these elements well to create a larger design \citep{bollini2017beautiful, hartmann2008towards}.

    \item \textit{\textbf{\textcolor{googlered}{\textsc{\bfseries \scshape Interest }}}}in terms of UIs, pertains to how eye-catching, inviting, and captivating an interface is. Interesting interfaces are capable of catching a user's attention and keeping it. An interesting interface may have UI elements such as unique fonts, mood-inducing colors, or sensational text headlines \cite{shneiderman2004designing}.

    \item \textit{\textbf{\textcolor{googlered}{\textsc{\bfseries \scshape Comfort }}}}pertains to the overall satisfaction and degree to which an interface makes a user feel at ease while in use \citep{dave2023understanding}. Designs and interfaces can increase comfort by maintaining a clean, straightforward, and low cognitive load design. A comfortable design may be more tonally consistent and use common fonts and colors.
\end{itemize}

    \begin{figure*}[t!]
    \centering
    \includegraphics[width=0.6\linewidth]{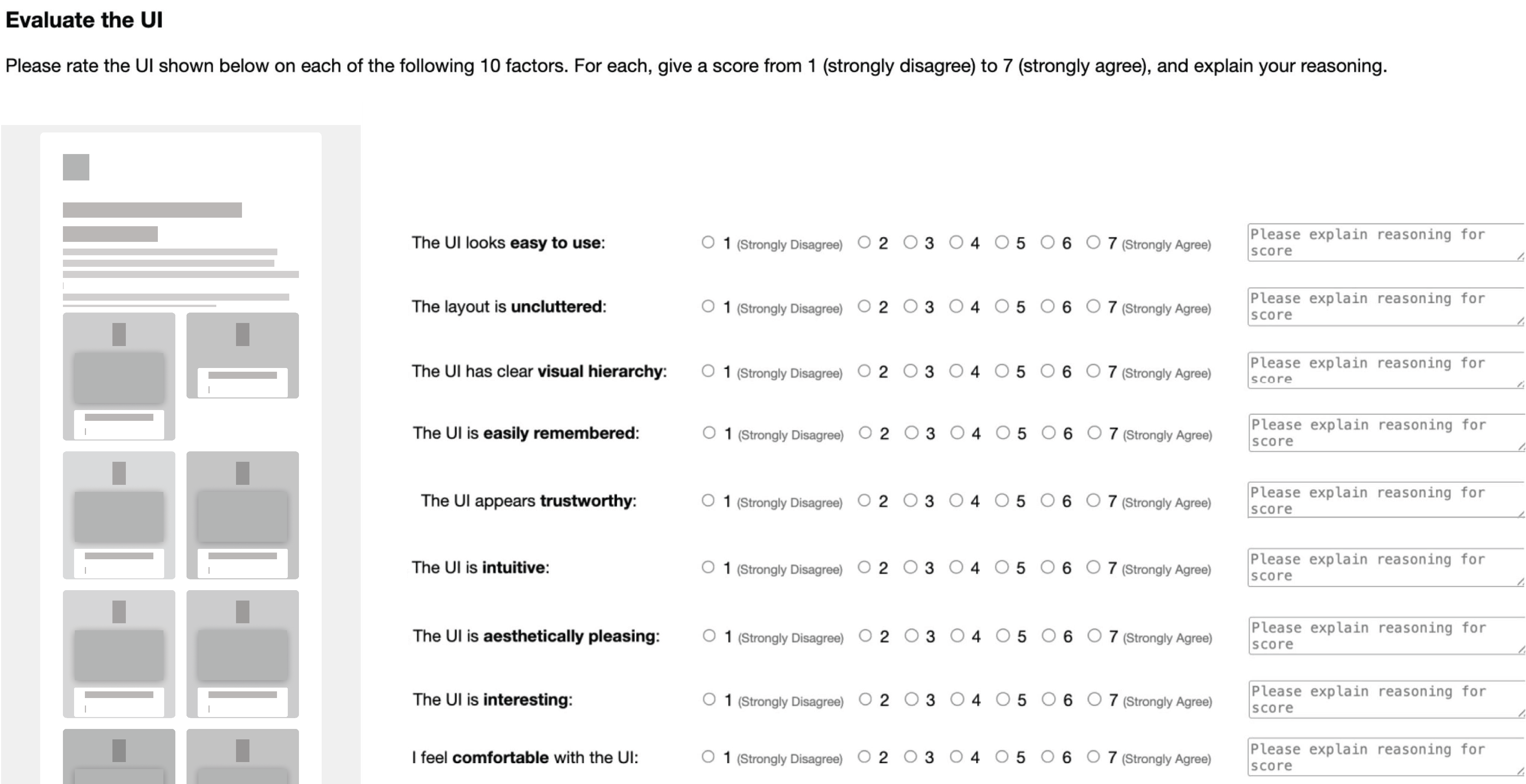}
    \caption{%
    {Example survey question shown to users. Consists of instructions on top, the UI on the left, and the nine factor, Likert scale questions on the right.
    } 
   }
    \label{fig:MTurkExample}
    \end{figure*}

\subsection{Participants and Procedure}
In total, we collected 15,000 responses from 500 participants using the online crowdsourcing platform Amazon's Mechanical Turk. %
To ensure the quality of the dataset, we filtered for participants who had a greater than 98\% approval rating and at least 500 responses to take part in our study. 
 In total, we received feedback on 30 UIs and we asked for 500 unique responses from each. From there, 
 we asked participants to evaluate each UI based on the nine UI Factors (\cref{table:Factors}). Each evaluation was expected to take about 45 seconds to a minute to complete. 
 In total, it cost \$75 to get 500 human evaluations for each UI. Before taxes and fees, the total cost of this part of the study was \$2,250. 

Each UI was presented as a standalone task to minimize order effects, and participants evaluated each UI in two stages.
First, participants responded to nine Likert scale questions, one corresponding to each of the UI Factors (see \cref{table:Factors}). 
Each question asked participants to indicate their level of agreement on a 7-point scale (1 = strongly disagree, 7 = strongly agree). 
To support their evaluation, participants were able to resize the UI during this stage, allowing them to view it while responding to the questions.
Next, participants were asked to provide a textual explanation describing the rationale behind their ratings. This qualitative input allowed for deeper insight into participants' reactions to each UI.

\subsection{Result of the Human Data Collection} \label{sec:ui-to-exp-result-of-data-collection}

We compiled a dataset containing each worker’s ID, the corresponding image ID, the UI’s domain, and the participant’s scores for each of the nine evaluation factors.
To ensure consistency across the benchmark, factor and domain names were standardized to align with the terminology used throughout the paper.
We applied several data pre-processing steps to ensure data quality. 
We excluded responses that were completed in less than 45 seconds, based on the rationale that a thoughtful engagement with the task requires at least that amount of time. 
Additionally, we removed responses that were incomplete or that contained duplicate scores submitted by the same user. Further, we removed data for screens that were not professionally made to maintain consistent data.
After the exclusion procedure, we are left with 9,296 human responses. 
\cref{table:LikertTable} presents the mean scores and standard deviations for each of the nine UI factors, as rated by human participants, split by domain.
The rightmost column shows the overall average score for each factor across all domains, while the bottom row reports the average score for each domain across all factors.
We then compared these human-generated ratings to scores produced by multimodal large language models (MLLMs) to evaluate alignment and divergence in trust-related perceptions.

\begin{table}[t!]
\centering
\small
\setlength{\tabcolsep}{1.0mm} 
\renewcommand{\arraystretch}{0.95}
\definecolor{nearA}{HTML}{D4EDDA} 
\definecolor{nearB}{HTML}{ECF8EC} 
\newcommand{\withinA}[1]{\cellcolor{nearA}#1}
\newcommand{\withinB}[1]{\cellcolor{nearB}#1}
\caption{Individual human and MLLM scores for all factors and domains. We report the average score over all UIs in the last column and over all factors in the last row. The closest MLLM score to the human is in bold. When the score is closer than $0.25$ ($0.5$), its background is green (light green). The gray text is the standard deviation of the scores from which the average is computed.}
\vspace{-2mm}
\begin{tabular}{@{}p{1.65cm}l ccc c@{}}
\toprule
& & \multicolumn{3}{c}{\sc Domains} \\
\cmidrule(){3-5}
\textbf{Factor} & \textbf{Evaluator} & \makecell{\textbf{Browsing} \\ \textbf{\& Disc.}} & \makecell{\textbf{Conf. \&} \\ \textbf{ Feedback}} & \makecell{\textbf{Comm. \&} \\ \textbf{Engage.}} & \textbf{All UIs} \\
\midrule
\multirow{4}{*}{\makecell{\textsc{\textcolor{googlegreen}{\textsc{\bfseries \scshape Ease of}}} \\ \textsc{\textcolor{googlegreen}{\textsc{\bfseries \scshape Use}}}}}
& Human & 6.20 \textcolor[HTML]{8C8C8C}{\scriptsize(1.08)} & 6.21 \textcolor[HTML]{8C8C8C}{\scriptsize(1.10)} & 6.21 \textcolor[HTML]{8C8C8C}{\scriptsize(1.07)} & 6.21 \textcolor[HTML]{8C8C8C}{\scriptsize(1.09)} \\
\cmidrule{2-6}
& Claude-3.5 & 7.00 \textcolor[HTML]{8C8C8C}{\scriptsize(0.00)} & 6.90 \textcolor[HTML]{8C8C8C}{\scriptsize(0.32)} & 6.78 \textcolor[HTML]{8C8C8C}{\scriptsize(0.67)} & 6.89 \textcolor[HTML]{8C8C8C}{\scriptsize(0.42)} \\
& GPT-4o & \withinB{\textbf{6.70} \textcolor[HTML]{8C8C8C}{\scriptsize(0.67)}} & \withinB{\textbf{6.50} \textcolor[HTML]{8C8C8C}{\scriptsize(0.97)}} & \withinA{\textbf{6.20} \textcolor[HTML]{8C8C8C}{\scriptsize(1.32)}} & \withinB{\textbf{6.47} \textcolor[HTML]{8C8C8C}{\scriptsize(1.01)}} \\
& Llama-3.2 & 6.80 \textcolor[HTML]{8C8C8C}{\scriptsize(0.42)} & \withinB{\textbf{6.50} \textcolor[HTML]{8C8C8C}{\scriptsize(0.53)}} & \withinA{6.40 \textcolor[HTML]{8C8C8C}{\scriptsize(0.70)}} & \withinB{6.57 \textcolor[HTML]{8C8C8C}{\scriptsize(0.57)}} \\
\midrule
\multirow{4}{*}{\textsc{\textcolor{googlegreen}{\textsc{\bfseries \scshape Clarity}}}}
& Human & 6.08 \textcolor[HTML]{8C8C8C}{\scriptsize(1.20)} & 6.14 \textcolor[HTML]{8C8C8C}{\scriptsize(1.21)} & 6.21 \textcolor[HTML]{8C8C8C}{\scriptsize(1.16)} & 6.14 \textcolor[HTML]{8C8C8C}{\scriptsize(1.19)} \\
\cmidrule{2-6}
& Claude-3.5 & \withinA{\textbf{6.00} \textcolor[HTML]{8C8C8C}{\scriptsize(0.50)}} & \withinA{5.90 \textcolor[HTML]{8C8C8C}{\scriptsize(0.57)}} & \withinA{\textbf{6.11} \textcolor[HTML]{8C8C8C}{\scriptsize(0.78)}} & \withinA{\textbf{6.00} \textcolor[HTML]{8C8C8C}{\scriptsize(0.61)}} \\
& GPT-4o & \withinB{5.60 \textcolor[HTML]{8C8C8C}{\scriptsize(1.71)}} & \withinA{\textbf{6.10} \textcolor[HTML]{8C8C8C}{\scriptsize(1.73)}} & 5.50 \textcolor[HTML]{8C8C8C}{\scriptsize(1.96)} & \withinB{5.73 \textcolor[HTML]{8C8C8C}{\scriptsize(1.76)}} \\
& Llama-3.2 & \withinB{6.50 \textcolor[HTML]{8C8C8C}{\scriptsize(0.85)}} & \withinB{6.50 \textcolor[HTML]{8C8C8C}{\scriptsize(0.71)}} & \withinB{6.60 \textcolor[HTML]{8C8C8C}{\scriptsize(0.70)}} & \withinB{6.53 \textcolor[HTML]{8C8C8C}{\scriptsize(0.73)}} \\
\midrule
\multirow{4}{*}{\makecell{\textsc{\textcolor{googlegreen}{\textsc{\bfseries \scshape Visual}}} \\ \textsc{\textcolor{googlegreen}{\textsc{\bfseries \scshape Hierarchy}}}}}
& Human & 6.19 \textcolor[HTML]{8C8C8C}{\scriptsize(1.08)} & 6.25 \textcolor[HTML]{8C8C8C}{\scriptsize(1.06)} & 6.26 \textcolor[HTML]{8C8C8C}{\scriptsize(1.04)} & 6.23 \textcolor[HTML]{8C8C8C}{\scriptsize(1.06)} \\
\cmidrule{2-6}
& Claude-3.5 & \withinB{6.56 \textcolor[HTML]{8C8C8C}{\scriptsize(0.53)}} & \withinA{6.10 \textcolor[HTML]{8C8C8C}{\scriptsize(0.57)}} & \withinA{6.44 \textcolor[HTML]{8C8C8C}{\scriptsize(0.53)}} & \withinA{6.37 \textcolor[HTML]{8C8C8C}{\scriptsize(0.56)}} \\
& GPT-4o & \withinA{\textbf{6.10} \textcolor[HTML]{8C8C8C}{\scriptsize(1.29)}} & \withinA{\textbf{6.30} \textcolor[HTML]{8C8C8C}{\scriptsize(1.34)}} & \withinA{\textbf{6.30} \textcolor[HTML]{8C8C8C}{\scriptsize(1.34)}} & \withinA{\textbf{6.23} \textcolor[HTML]{8C8C8C}{\scriptsize(1.28)}} \\
& Llama-3.2 & 6.80 \textcolor[HTML]{8C8C8C}{\scriptsize(0.42)} & \withinA{6.50 \textcolor[HTML]{8C8C8C}{\scriptsize(0.53)}} & 6.80 \textcolor[HTML]{8C8C8C}{\scriptsize(0.42)} & \withinB{6.70 \textcolor[HTML]{8C8C8C}{\scriptsize(0.47)}} \\
\midrule
\multirow{4}{*}{\textsc{\textcolor{googleblue}{{\bfseries \scshape Memorable}}}}
& Human & 5.79 \textcolor[HTML]{8C8C8C}{\scriptsize(0.97)} & 5.79 \textcolor[HTML]{8C8C8C}{\scriptsize(0.97)} & 5.86 \textcolor[HTML]{8C8C8C}{\scriptsize(0.91)} & 5.81 \textcolor[HTML]{8C8C8C}{\scriptsize(0.95)} \\
\cmidrule{2-6}
& Claude-3.5 & \withinB{5.44 \textcolor[HTML]{8C8C8C}{\scriptsize(0.73)}} & 5.10 \textcolor[HTML]{8C8C8C}{\scriptsize(0.57)} & \withinA{6.00 \textcolor[HTML]{8C8C8C}{\scriptsize(0.87)}} & \withinB{5.51 \textcolor[HTML]{8C8C8C}{\scriptsize(0.79)}} \\
& GPT-4o & \withinA{\textbf{5.70} \textcolor[HTML]{8C8C8C}{\scriptsize(0.67)}} & \withinA{\textbf{5.70} \textcolor[HTML]{8C8C8C}{\scriptsize(0.67)}} & \withinA{\textbf{5.80} \textcolor[HTML]{8C8C8C}{\scriptsize(0.42)}} & \withinA{\textbf{5.73} \textcolor[HTML]{8C8C8C}{\scriptsize(0.58)}} \\
& Llama-3.2 & \withinB{5.30 \textcolor[HTML]{8C8C8C}{\scriptsize(0.48)}} & 5.20 \textcolor[HTML]{8C8C8C}{\scriptsize(0.42)} & \withinA{\textbf{5.80} \textcolor[HTML]{8C8C8C}{\scriptsize(0.92)}} & \withinB{5.43 \textcolor[HTML]{8C8C8C}{\scriptsize(0.68)}} \\
\midrule
\multirow{4}{*}{\textsc{\textcolor{googleblue}{{\bfseries \scshape Trust}}}}
& Human & 6.31 \textcolor[HTML]{8C8C8C}{\scriptsize(1.02)} & 6.41 \textcolor[HTML]{8C8C8C}{\scriptsize(0.99)} & 6.29 \textcolor[HTML]{8C8C8C}{\scriptsize(1.01)} & 6.34 \textcolor[HTML]{8C8C8C}{\scriptsize(1.01)} \\
\cmidrule{2-6}
& Claude-3.5 & \withinA{6.56 \textcolor[HTML]{8C8C8C}{\scriptsize(0.53)}} & \withinB{6.70 \textcolor[HTML]{8C8C8C}{\scriptsize(0.48)}} & \withinB{6.00 \textcolor[HTML]{8C8C8C}{\scriptsize(0.50)}} & \withinA{6.42 \textcolor[HTML]{8C8C8C}{\scriptsize(0.57)}} \\
& GPT-4o & \withinB{6.70 \textcolor[HTML]{8C8C8C}{\scriptsize(0.67)}} & \withinA{\textbf{6.60} \textcolor[HTML]{8C8C8C}{\scriptsize(0.70)}} & \withinA{\textbf{6.30} \textcolor[HTML]{8C8C8C}{\scriptsize(0.67)}} & \withinA{6.53 \textcolor[HTML]{8C8C8C}{\scriptsize(0.68)}} \\
& Llama-3.2 & \withinA{\textbf{6.50} \textcolor[HTML]{8C8C8C}{\scriptsize(0.53)}} & \withinA{\textbf{6.60} \textcolor[HTML]{8C8C8C}{\scriptsize(0.70)}} & \withinB{6.00 \textcolor[HTML]{8C8C8C}{\scriptsize(0.82)}} & \withinA{\textbf{6.37} \textcolor[HTML]{8C8C8C}{\scriptsize(0.72)}} \\
\midrule
\multirow{4}{*}{\textsc{\textcolor{googleblue}{{\bfseries \scshape Intuitive}}}}
& Human & 6.22 \textcolor[HTML]{8C8C8C}{\scriptsize(1.07)} & 6.31 \textcolor[HTML]{8C8C8C}{\scriptsize(1.04)} & 6.25 \textcolor[HTML]{8C8C8C}{\scriptsize(1.03)} & 6.26 \textcolor[HTML]{8C8C8C}{\scriptsize(1.05)} \\
\cmidrule{2-6}
& Claude-3.5 & 7.00 \textcolor[HTML]{8C8C8C}{\scriptsize(0.00)} & 6.90 \textcolor[HTML]{8C8C8C}{\scriptsize(0.32)} & \withinA{\textbf{6.22} \textcolor[HTML]{8C8C8C}{\scriptsize(0.67)}} & \withinB{6.71 \textcolor[HTML]{8C8C8C}{\scriptsize(0.53)}} \\
& GPT-4o & \withinA{\textbf{6.20} \textcolor[HTML]{8C8C8C}{\scriptsize(0.92)}} & \withinA{6.40 \textcolor[HTML]{8C8C8C}{\scriptsize(0.97)}} & \withinB{5.90 \textcolor[HTML]{8C8C8C}{\scriptsize(1.20)}} & \withinA{6.17 \textcolor[HTML]{8C8C8C}{\scriptsize(1.02)}} \\
& Llama-3.2 & \withinB{6.60 \textcolor[HTML]{8C8C8C}{\scriptsize(0.52)}} & \withinA{\textbf{6.30} \textcolor[HTML]{8C8C8C}{\scriptsize(0.48)}} & \withinA{6.00 \textcolor[HTML]{8C8C8C}{\scriptsize(0.47)}} & \withinA{\textbf{6.30} \textcolor[HTML]{8C8C8C}{\scriptsize(0.53)}} \\
\midrule
\multirow{4}{*}{\makecell{\textsc{\textcolor{googlered}{{\bfseries \scshape Aesthetic}}} \\ \textsc{\textcolor{googlered}{{\bfseries \scshape Pleasure}}}}}
& Human & 5.86 \textcolor[HTML]{8C8C8C}{\scriptsize(0.91)} & 5.85 \textcolor[HTML]{8C8C8C}{\scriptsize(0.93)} & 6.06 \textcolor[HTML]{8C8C8C}{\scriptsize(0.98)} & 5.92 \textcolor[HTML]{8C8C8C}{\scriptsize(0.94)} \\
\cmidrule{2-6}
& Claude-3.5 & \withinA{5.78 \textcolor[HTML]{8C8C8C}{\scriptsize(0.44)}} & \withinB{\textbf{5.50} \textcolor[HTML]{8C8C8C}{\scriptsize(1.08)}} & \withinB{6.33 \textcolor[HTML]{8C8C8C}{\scriptsize(0.71)}} & \withinA{\textbf{5.87} \textcolor[HTML]{8C8C8C}{\scriptsize(0.85)}} \\
& GPT-4o & 5.10 \textcolor[HTML]{8C8C8C}{\scriptsize(1.45)} & 5.30 \textcolor[HTML]{8C8C8C}{\scriptsize(1.57)} & \withinB{5.80 \textcolor[HTML]{8C8C8C}{\scriptsize(1.23)}} & 5.40 \textcolor[HTML]{8C8C8C}{\scriptsize(1.40)} \\
& Llama-3.2 & \withinA{\textbf{5.80} \textcolor[HTML]{8C8C8C}{\scriptsize(0.63)}} & 5.00 \textcolor[HTML]{8C8C8C}{\scriptsize(1.05)} & \withinA{\textbf{6.30} \textcolor[HTML]{8C8C8C}{\scriptsize(0.82)}} & \withinA{5.70 \textcolor[HTML]{8C8C8C}{\scriptsize(0.99)}} \\
\midrule
\multirow{4}{*}{\textsc{\textcolor{googlered}{{\bfseries \scshape Interesting}}}}
& Human & 5.79 \textcolor[HTML]{8C8C8C}{\scriptsize(0.99)} & 5.70 \textcolor[HTML]{8C8C8C}{\scriptsize(1.08)} & 5.96 \textcolor[HTML]{8C8C8C}{\scriptsize(0.97)} & 5.82 \textcolor[HTML]{8C8C8C}{\scriptsize(1.02)} \\
\cmidrule{2-6}
& Claude-3.5 & \textbf{5.22} \textcolor[HTML]{8C8C8C}{\scriptsize(0.44)} & \textbf{4.40} \textcolor[HTML]{8C8C8C}{\scriptsize(0.70)} & \withinB{\textbf{5.67} \textcolor[HTML]{8C8C8C}{\scriptsize(0.50)}} & \textbf{5.10} \textcolor[HTML]{8C8C8C}{\scriptsize(0.77)} \\
& GPT-4o & 4.50 \textcolor[HTML]{8C8C8C}{\scriptsize(0.71)} & 4.10 \textcolor[HTML]{8C8C8C}{\scriptsize(0.57)} & \withinB{5.60 \textcolor[HTML]{8C8C8C}{\scriptsize(0.84)}} & 4.73 \textcolor[HTML]{8C8C8C}{\scriptsize(0.94)} \\
& Llama-3.2 & 4.60 \textcolor[HTML]{8C8C8C}{\scriptsize(0.70)} & 3.90 \textcolor[HTML]{8C8C8C}{\scriptsize(0.99)} & 5.00 \textcolor[HTML]{8C8C8C}{\scriptsize(0.94)} & 4.50 \textcolor[HTML]{8C8C8C}{\scriptsize(0.97)} \\
\midrule
\multirow{4}{*}{\textsc{\textcolor{googlered}{{\bfseries \scshape Comfort}}}}
& Human & 6.23 \textcolor[HTML]{8C8C8C}{\scriptsize(1.03)} & 6.30 \textcolor[HTML]{8C8C8C}{\scriptsize(1.03)} & 6.28 \textcolor[HTML]{8C8C8C}{\scriptsize(1.00)} & 6.27 \textcolor[HTML]{8C8C8C}{\scriptsize(1.02)} \\
\cmidrule{2-6}
& Claude-3.5 & \withinA{6.11 \textcolor[HTML]{8C8C8C}{\scriptsize(0.33)}} & \withinA{\textbf{6.40} \textcolor[HTML]{8C8C8C}{\scriptsize(0.70)}} & \withinB{\textbf{6.00} \textcolor[HTML]{8C8C8C}{\scriptsize(0.00)}} & \withinA{\textbf{6.17} \textcolor[HTML]{8C8C8C}{\scriptsize(0.48)}} \\
& GPT-4o & 5.70 \textcolor[HTML]{8C8C8C}{\scriptsize(0.67)} & \withinB{5.80 \textcolor[HTML]{8C8C8C}{\scriptsize(1.14)}} & 5.70 \textcolor[HTML]{8C8C8C}{\scriptsize(0.95)} & 5.73 \textcolor[HTML]{8C8C8C}{\scriptsize(0.91)} \\
& Llama-3.2 & \withinA{\textbf{6.30} \textcolor[HTML]{8C8C8C}{\scriptsize(0.67)}} & \withinB{5.90 \textcolor[HTML]{8C8C8C}{\scriptsize(0.57)}} & \withinB{\textbf{6.00} \textcolor[HTML]{8C8C8C}{\scriptsize(0.47)}} & \withinA{6.07 \textcolor[HTML]{8C8C8C}{\scriptsize(0.58)}} \\
\midrule
\multirow{4}{*}{\textsc{\bfseries \scshape All}}
& Human & 6.06 \textcolor[HTML]{8C8C8C}{\scriptsize(1.06)} & 6.10 \textcolor[HTML]{8C8C8C}{\scriptsize(1.08)} & 6.13 \textcolor[HTML]{8C8C8C}{\scriptsize(1.03)} & 6.10 \textcolor[HTML]{8C8C8C}{\scriptsize(1.06)} \\
\cmidrule{2-6}
& Claude-3.5 & \withinA{6.14 \textcolor[HTML]{8C8C8C}{\scriptsize(0.73)}} & \withinA{\textbf{5.97} \textcolor[HTML]{8C8C8C}{\scriptsize(0.97)}} & \withinA{\textbf{6.14} \textcolor[HTML]{8C8C8C}{\scriptsize(0.65)}} & \withinA{\textbf{6.08} \textcolor[HTML]{8C8C8C}{\scriptsize(0.80)}} \\
& GPT-4o & \withinB{5.70 \textcolor[HTML]{8C8C8C}{\scriptsize(1.21)}} & \withinB{5.75 \textcolor[HTML]{8C8C8C}{\scriptsize(1.31)}} & \withinB{5.80 \textcolor[HTML]{8C8C8C}{\scriptsize(1.15)}} & \withinB{5.75 \textcolor[HTML]{8C8C8C}{\scriptsize(1.22)}} \\
& Llama-3.2 & \withinA{\textbf{6.07} \textcolor[HTML]{8C8C8C}{\scriptsize(0.91)}} & \withinB{5.80 \textcolor[HTML]{8C8C8C}{\scriptsize(1.07)}} & \withinA{6.07 \textcolor[HTML]{8C8C8C}{\scriptsize(0.82)}} & \withinA{5.98 \textcolor[HTML]{8C8C8C}{\scriptsize(0.95)}} \\
\bottomrule
\end{tabular}
\label{table:LikertTable}
\end{table}

\section{Evaluation of MLLM Judges for UIs}
\label{sec:exp}

We evaluate the efficacy of MLLMs in assessing UIs by comparing their judgments to human evaluations collected through a crowdsourcing study (\cref{sec:ui-to-exp}). 
Our analysis proceeds in two parts. 
In \cref{sec:exp-score-pred}, we examine how well MLLMs predict absolute human scores (prompt shown in \cref{fig:prompt-UI1}), assessing both the accuracy of these predictions and their utility for estimating user preferences. 
In \cref{sec:exp-pairwise-comparison}, we analyze pairwise comparisons of UIs based on human data, and evaluate how closely MLLM predictions of these comparisons align with human preferences (prompt shown in \cref{fig:prompt-pairwise}).
The analysis is conducted over all UI factors in \cref{table:Factors}. To avoid bias, and ensure generality and robustness of our findings, we leverage three diverse MLLMs, including GPT-4o, Claude 3.5 Sonnet, and Llama‑3.2‑11B‑Vision.

\subsection{Task 1: Absolute Score Prediction}
\label{sec:exp-score-pred}

We conduct three studies evaluating how closely MLLMs’ responses to UIs resemble those of humans. 
In \cref{sec:individual scores}, we first analyze MLLM and human scores separately. Results suggest that MLLMs are mostly well aligned to humans, with a few factors where they may significantly under or over predict human scores. 
In \cref{sec:score prediction}, we analyze prediction errors of MLLMs using four metrics: MSE, MAE, accuracy, and $\pm 1$-accuracy.
Results suggest that MLLMs well-predict human responses, with Claude being the best judge overall. MLLMs on average overestimated factors like \textcolor{googlegreen}{\scshape Ease of Use}. While consistently underestimating factors like \textcolor{googlered}{\scshape Aesthetic Pleasure}, \textcolor{googlered}{\scshape Interesting}, and \textcolor{googlered}{\scshape Comfort}. Meanwhile, models showed they could reasonably predict scores (within 0.50 Likert points) for factors such as \textcolor{googlegreen}{\scshape Clarity}, \textcolor{googlegreen}{\scshape Visual Hierarchy}, \textcolor{googleblue}{\scshape Memorable}, and \textcolor{googleblue}{\scshape Trustworthy}, \textcolor{googleblue}{\scshape Intuitive}. 
Finally in \cref{sec:ranking prediction}, we analyze how well MLLM scores predict the ranking of UIs by humans. The factor scan be well predicted, and the best judge overall is Llama. 
To reduce the randomness in our analysis due to random MLLM scores, we evaluate each UI and factor by each MLLM $10$ times and then take the median score. Our MLLM prompt is reported in \cref{fig:prompt-UI1} in the Appendix.

\subsubsection{Individual Scores}
\label{sec:individual scores}

We start by comparing individual human and MLLM scores. 
For each UI and factor, we report averaged human scores and median MLLM scores.
These averages represent an average perception of a given UI in a given factor. 
We also included average scores from humans and MLLMs aggregated by the three domains and all UIs, see~\cref{table:LikertTable}.

We make several observations. First, MLLM judges align reasonably well with humans. As an example, note the average human and MLLM scores over all factors and UIs in \cref{table:LikertTable} (last four rows and last column). 
The human score is $6.10$, whereas Claude predicts $6.08$, GPT-4o predicts $5.75$, and Llama predicts $5.98$. Therefore, all MLLM scores are within $0.35$ of the human scores. Since this is significantly less than the range of our $7$-point Likert scale, we conclude that MLLMs reasonably mimic human scores. 
Second, MLLM scores are often closer to each other than those of humans, as indicated by smaller standard deviations of the scores in \cref{table:LikertTable}. We note this specifically for Claude and Llama. GPT-4o has sometimes a larger standard deviation and sometimes smaller.

\emph{When considering individual factors in \cref{table:LikertTable}, MLLMs tend to perform well on cognitive and perceptual factors.} For instance, the human score for \textcolor{googleblue}{\scshape Trustworthy} is $6.34$, whereas Claude predicts $6.42$, GPT-4o predicts $6.53$, and Llama predicts $6.37$. This is the only factor where all MLLM scores are within $0.25$ of the human score. MLLMs also perform well on \textcolor{googlegreen}{\scshape Visual Hierarchy}, where all MLLM scores are within $0.5$ of the human scores. The human score is $6.23$, while Claude predicts $6.37$, GPT-4o predicts $6.23$, and Llama predicts $6.70$. Overall, MLLMs likely approximate these scores effectively because the factors have concrete cues in the UI. These factors are correlated with layout, reassuring designs and phrases, and spacing; all of which MLLMs can gather from the UI screenshots.

\emph{MLLMs struggle with some emotional factors, such as \textcolor{googlered}{\scshape Interesting}.} Specifically, all MLLM scores of this factor underestimate human scores. The human score is $5.82$, whereas Claude predicts $5.10$, GPT-4o predicts $4.73$, and Llama predicts $4.50$. The MLLMs likely struggle with this and other factors like \textcolor{googlegreen}{\scshape Ease of Use} because they are more difficult to quantify for MLLMs. Factors like ease of use are highly tied a users technical literacy and cannot be inferred from visual cues alone \citep{muzaffar2011usability,hinton2012using}.
\begin{table*}[h!]
\centering
\small
\setlength{\tabcolsep}{.35em}
\renewcommand{\arraystretch}{1.0}
\caption{We report seven error metrics for all factors and MLLM judges. Four metrics measure the ability to predict scores (MSE, MAE, accuracy, and $\pm 1$-accuracy) and three measure their utility for ranking (Pearson correlation coefficient, Spearman's $\rho$, and Kendall's $\tau$). The last row contains metric values for all factors jointly. Higher correlation coefficients than $0.4$ are reported in bold.}
\begin{tabular}{@{}p{1.8cm}p{2.2cm}lccccccc@{}}
\toprule
& \textbf{Factor} & \textbf{Model} & \makecell{\textbf{MSE} $\downarrow$} & \makecell{\textbf{MAE} $\downarrow$} & 
\textbf{Acc.} $\uparrow$ & 
\textbf{Acc. $\pm1$} & 
\makecell{\textbf{Pearson} $\uparrow$} & \makecell{\textbf{Spearman} $\uparrow$}  & \makecell{\textbf{Kendall} $\uparrow$} \\
\midrule
\multirow{9}{*}{\textcolor{googlegreen}{\textsc{\sc \bfseries Cognitive}}} & \multirow{3}{*}{{\makecell[l]{\textsc{Ease of} \\ \textsc{Use}}}} & Claude-3.5 & 0.83 & 0.87 & 0.42 & 0.76 & \makecell{-0.05 \ \textcolor[HTML]{8C8C8C}{\scriptsize(0.820)}} & \makecell{-0.32 \ \textcolor[HTML]{8C8C8C}{\scriptsize(0.152)}} & \makecell{-0.25 \ \textcolor[HTML]{8C8C8C}{\scriptsize(0.163)}} \\
 & & GPT-4o & 0.56 & 0.72 & 0.42 & 0.77 & \makecell{0.19 \ \textcolor[HTML]{8C8C8C}{\scriptsize(0.380)}} & \makecell{0.14 \ \textcolor[HTML]{8C8C8C}{\scriptsize(0.529)}} & \makecell{0.10 \ \textcolor[HTML]{8C8C8C}{\scriptsize(0.514)}} \\
 & & Llama-3.2 & 0.37 & 0.54 & 0.40 & 0.79 & \makecell{0.23 \ \textcolor[HTML]{8C8C8C}{\scriptsize(0.283)}} & \makecell{0.20 \ \textcolor[HTML]{8C8C8C}{\scriptsize(0.348)}} & \makecell{0.13 \ \textcolor[HTML]{8C8C8C}{\scriptsize(0.396)}} \\
\cmidrule(lr){2-10}
 & \multirow{3}{*}{\textsc{Clarity}} & Claude-3.5 & 0.10 & 0.22 & 0.33 & 0.81 & \makecell{0.21 \ \textcolor[HTML]{8C8C8C}{\scriptsize(0.341)}} & \makecell{0.22 \ \textcolor[HTML]{8C8C8C}{\scriptsize(0.324)}} & \makecell{0.18 \ \textcolor[HTML]{8C8C8C}{\scriptsize(0.292)}} \\
 & & GPT-4o & 0.48 & 0.63 & 0.38 & 0.69 & \makecell{-0.07 \ \textcolor[HTML]{8C8C8C}{\scriptsize(0.744)}} & \makecell{0.06 \ \textcolor[HTML]{8C8C8C}{\scriptsize(0.788)}} & \makecell{0.05 \ \textcolor[HTML]{8C8C8C}{\scriptsize(0.764)}} \\
 & & Llama-3.2 & 0.45 & 0.58 & 0.38 & 0.75 & \makecell{0.34 \ \textcolor[HTML]{8C8C8C}{\scriptsize(0.100)}} & \makecell{0.27 \ \textcolor[HTML]{8C8C8C}{\scriptsize(0.198)}} & \makecell{0.19 \ \textcolor[HTML]{8C8C8C}{\scriptsize(0.202)}} \\
\cmidrule(lr){2-10}
 & \multirow{3}{*}{{\makecell[l]{\textsc{Visual} \\ \textsc{Hierarchy}}}} & Claude-3.5 & 0.26 & 0.41 & 0.37 & 0.78 & \makecell{-0.14 \ \textcolor[HTML]{8C8C8C}{\scriptsize(0.535)}} & \makecell{-0.06 \ \textcolor[HTML]{8C8C8C}{\scriptsize(0.779)}} & \makecell{-0.06 \ \textcolor[HTML]{8C8C8C}{\scriptsize(0.710)}} \\
 & & GPT-4o & 0.43 & 0.60 & 0.40 & 0.75 & \makecell{0.02 \ \textcolor[HTML]{8C8C8C}{\scriptsize(0.929)}} & \makecell{-0.04 \ \textcolor[HTML]{8C8C8C}{\scriptsize(0.845)}} & \makecell{-0.05 \ \textcolor[HTML]{8C8C8C}{\scriptsize(0.744)}} \\
 & & Llama-3.2 & 0.38 & 0.56 & 0.42 & 0.78 & \makecell{0.38 \ \textcolor[HTML]{8C8C8C}{\scriptsize(0.068)}} & \makecell{\textbf{0.54} \ \textcolor[HTML]{8C8C8C}{\scriptsize(0.007)}} & \makecell{\textbf{0.41} \ \textcolor[HTML]{8C8C8C}{\scriptsize(0.006)}} \\
\midrule
\multirow{9}{*}{\textcolor{googleblue}{\textsc{\sc \bfseries Perceptual}}} & 
\multirow{3}{*}{\textsc{Memorable}} & Claude-3.5 & 0.36 & 0.53 & 0.28 & 0.72 & \makecell{\textbf{0.59} \ \textcolor[HTML]{8C8C8C}{\scriptsize(0.004)}} & \makecell{0.39 \ \textcolor[HTML]{8C8C8C}{\scriptsize(0.069)}} & \makecell{0.27 \ \textcolor[HTML]{8C8C8C}{\scriptsize(0.092)}} \\
 & & GPT-4o & 0.08 & 0.23 & 0.37 & 0.67 & \makecell{\textbf{0.41} \ \textcolor[HTML]{8C8C8C}{\scriptsize(0.046)}} & \makecell{0.32 \ \textcolor[HTML]{8C8C8C}{\scriptsize(0.131)}} & \makecell{0.24 \ \textcolor[HTML]{8C8C8C}{\scriptsize(0.115)}} \\
 & & Llama-3.2 & 0.34 & 0.52 & 0.30 & 0.75 & \makecell{\textbf{0.47} \ \textcolor[HTML]{8C8C8C}{\scriptsize(0.019)}} & \makecell{\textbf{0.46} \ \textcolor[HTML]{8C8C8C}{\scriptsize(0.023)}} & \makecell{\textbf{0.36} \ \textcolor[HTML]{8C8C8C}{\scriptsize(0.024)}} \\
\cmidrule(lr){2-10}
 & \multirow{3}{*}{\textsc{Trustworthy}} & Claude-3.5 & 0.40 & 0.58 & 0.44 & 0.80 & \makecell{0.10 \ \textcolor[HTML]{8C8C8C}{\scriptsize(0.652)}} & \makecell{0.13 \ \textcolor[HTML]{8C8C8C}{\scriptsize(0.573)}} & \makecell{0.08 \ \textcolor[HTML]{8C8C8C}{\scriptsize(0.638)}} \\
 & & GPT-4o & 0.27 & 0.45 & 0.44 & 0.82 & \makecell{0.21 \ \textcolor[HTML]{8C8C8C}{\scriptsize(0.320)}} & \makecell{0.00 \ \textcolor[HTML]{8C8C8C}{\scriptsize(0.995)}} & \makecell{-0.01 \ \textcolor[HTML]{8C8C8C}{\scriptsize(0.940)}} \\
 & & Llama-3.2 & 0.32 & 0.48 & 0.43 & 0.84 & \makecell{0.40 \ \textcolor[HTML]{8C8C8C}{\scriptsize(0.052)}} & \makecell{0.33 \ \textcolor[HTML]{8C8C8C}{\scriptsize(0.116)}} & \makecell{0.24 \ \textcolor[HTML]{8C8C8C}{\scriptsize(0.135)}} \\
\cmidrule(lr){2-10}
 & \multirow{3}{*}{\textsc{Intuitive}} & Claude-3.5 & 0.58 & 0.71 & 0.43 & 0.78 & \makecell{-0.02 \ \textcolor[HTML]{8C8C8C}{\scriptsize(0.944)}} & \makecell{0.06 \ \textcolor[HTML]{8C8C8C}{\scriptsize(0.794)}} & \makecell{0.03 \ \textcolor[HTML]{8C8C8C}{\scriptsize(0.843)}} \\
 & & GPT-4o & 0.20 & 0.38 & 0.40 & 0.79 & \makecell{0.24 \ \textcolor[HTML]{8C8C8C}{\scriptsize(0.254)}} & \makecell{0.22 \ \textcolor[HTML]{8C8C8C}{\scriptsize(0.296)}} & \makecell{0.17 \ \textcolor[HTML]{8C8C8C}{\scriptsize(0.270)}} \\
 & & Llama-3.2 & 0.13 & 0.28 & 0.36 & 0.81 & \makecell{0.02 \ \textcolor[HTML]{8C8C8C}{\scriptsize(0.919)}} & \makecell{-0.18 \ \textcolor[HTML]{8C8C8C}{\scriptsize(0.410)}} & \makecell{-0.14 \ \textcolor[HTML]{8C8C8C}{\scriptsize(0.340)}} \\
\midrule
\multirow{9}{*}{\textcolor{googlered}{\textsc{\sc \bfseries Emotional}}} & 
\multirow{3}{*}{{\makecell[l]{\textsc{Aesthetic} \\ \textsc{Pleasure}}}} & Claude-3.5 & 0.16 & 0.31 & 0.49 & 0.85 & \makecell{\textbf{0.62} \ \textcolor[HTML]{8C8C8C}{\scriptsize(<0.001)}} & \makecell{\textbf{0.62} \ \textcolor[HTML]{8C8C8C}{\scriptsize(<0.001)}} & \makecell{\textbf{0.49} \ \textcolor[HTML]{8C8C8C}{\scriptsize(<0.001)}} \\
 & & GPT-4o & 0.24 & 0.38 & 0.41 & 0.76 & \makecell{0.21 \ \textcolor[HTML]{8C8C8C}{\scriptsize(0.335)}} & \makecell{0.19 \ \textcolor[HTML]{8C8C8C}{\scriptsize(0.384)}} & \makecell{0.10 \ \textcolor[HTML]{8C8C8C}{\scriptsize(0.513)}} \\
 & & Llama-3.2 & 0.38 & 0.47 & 0.38 & 0.76 & \makecell{\textbf{0.51} \ \textcolor[HTML]{8C8C8C}{\scriptsize(0.012)}} & \makecell{\textbf{0.55} \ \textcolor[HTML]{8C8C8C}{\scriptsize(0.005)}} & \makecell{\textbf{0.42} \ \textcolor[HTML]{8C8C8C}{\scriptsize(0.005)}} \\
\cmidrule(lr){2-10}
 & \multirow{3}{*}{\textsc{Interesting}} & Claude-3.5 & 0.51 & 0.62 & 0.26 & 0.62 & \makecell{\textbf{0.85} \ \textcolor[HTML]{8C8C8C}{\scriptsize(<0.001)}} & \makecell{\textbf{0.82} \ \textcolor[HTML]{8C8C8C}{\scriptsize(<0.001)}} & \makecell{\textbf{0.69} \ \textcolor[HTML]{8C8C8C}{\scriptsize(<0.001)}} \\
 & & GPT-4o & 1.38 & 1.09 & 0.14 & 0.43 & \makecell{\textbf{0.63} \ \textcolor[HTML]{8C8C8C}{\scriptsize(<0.001)}} & \makecell{\textbf{0.59} \ \textcolor[HTML]{8C8C8C}{\scriptsize(0.002)}} & \makecell{\textbf{0.47} \ \textcolor[HTML]{8C8C8C}{\scriptsize(0.002)}} \\
 & & Llama-3.2 & 1.78 & 1.19 & 0.16 & 0.37 & \makecell{\textbf{0.61} \ \textcolor[HTML]{8C8C8C}{\scriptsize(0.002)}} & \makecell{\textbf{0.60} \ \textcolor[HTML]{8C8C8C}{\scriptsize(0.002)}} & \makecell{\textbf{0.44} \ \textcolor[HTML]{8C8C8C}{\scriptsize(0.003)}} \\
\cmidrule(lr){2-10}
 & \multirow{3}{*}{\textsc{Comfort}} & Claude-3.5 & 0.17 & 0.29 & 0.36 & 0.85 & \makecell{-0.12 \ \textcolor[HTML]{8C8C8C}{\scriptsize(0.583)}} & \makecell{-0.10 \ \textcolor[HTML]{8C8C8C}{\scriptsize(0.664)}} & \makecell{-0.05 \ \textcolor[HTML]{8C8C8C}{\scriptsize(0.743)}} \\
 & & GPT-4o & 0.08 & 0.24 & 0.35 & 0.78 & \makecell{-0.29 \ \textcolor[HTML]{8C8C8C}{\scriptsize(0.171)}} & \makecell{-0.32 \ \textcolor[HTML]{8C8C8C}{\scriptsize(0.124)}} & \makecell{-0.25 \ \textcolor[HTML]{8C8C8C}{\scriptsize(0.096)}} \\
 & & Llama-3.2 & 0.13 & 0.25 & 0.34 & 0.82 & \makecell{0.22 \ \textcolor[HTML]{8C8C8C}{\scriptsize(0.310)}} & \makecell{0.24 \ \textcolor[HTML]{8C8C8C}{\scriptsize(0.251)}} & \makecell{0.18 \ \textcolor[HTML]{8C8C8C}{\scriptsize(0.224)}} \\
\midrule
 & \multirow{3}{*}{\textsc{All}} & Claude-3.5 & 0.37 & 0.51 & 0.38 & 0.77 & \makecell{\textbf{0.69} \ \textcolor[HTML]{8C8C8C}{\scriptsize(<0.001)}} & \makecell{\textbf{0.58} \ \textcolor[HTML]{8C8C8C}{\scriptsize(<0.001)}} & \makecell{\textbf{0.44} \ \textcolor[HTML]{8C8C8C}{\scriptsize(<0.001)}} \\
 & & GPT-4o & 0.41 & 0.52 & 0.37 & 0.72 & \makecell{\textbf{0.70} \ \textcolor[HTML]{8C8C8C}{\scriptsize(<0.001)}} & \makecell{\textbf{0.60} \ \textcolor[HTML]{8C8C8C}{\scriptsize(<0.001)}} & \makecell{\textbf{0.43} \ \textcolor[HTML]{8C8C8C}{\scriptsize(<0.001)}} \\
 & & Llama-3.2 & 0.48 & 0.54 & 0.35 & 0.74 & \makecell{\textbf{0.73} \ \textcolor[HTML]{8C8C8C}{\scriptsize(<0.001)}} & \makecell{\textbf{0.63} \ \textcolor[HTML]{8C8C8C}{\scriptsize(<0.001)}} & \makecell{\textbf{0.47} \ \textcolor[HTML]{8C8C8C}{\scriptsize(<0.001)}} \\
\bottomrule
\end{tabular}
\label{Table:FactorsScores}
\end{table*}

\subsubsection{Score Prediction}
\label{sec:score prediction}

Next we analyze how well MLLMs predict human scores. We consider four error metrics: mean squared error (MSE), mean absolute error (MAE), accuracy, and $\pm 1$-accuracy. All metrics are computed on average human and median MLLM scores, for each UI and factor, as in \cref{sec:individual scores}. The \emph{MSE} is the mean of squared differences between MLLM and human scores, over all UIs and for a given factor. The \emph{MAE} is the mean of absolute differences between MLLM and human scores, over all UIs and for a given factor. The \emph{accuracy} is the fraction of MLLM scores that matches rounded human scores, over all UIs and for a given factor. The \emph{$\pm 1$-accuracy} is the fraction of MLLM scores that matches rounded human scores within $\pm 1$, over all UIs and for a given factor. The four metrics provide different views on our results. The MSE and MAE measure mean deviations in the MLLM and human scores, with the former penalizing larger deviations more. The accuracy measures the probability that the MLLM score matches the human score. The $\pm 1$-accuracy measures the probability of a sufficiently close match. All metrics are reported in \cref{Table:FactorsScores}.

MSE and MAE of some factors are low, which indicates predictability. For instance, all MSEs of \textcolor{googlered}{\scshape Comfort} are lower than $0.2$ and the lowest of all factors, and all its MAEs are lower than $0.3$ and the lowest of all factors. On the other hand, MSE and MAE of some factors are high, which indicates unpredictability. For instance, both MSEs and MAEs of \textcolor{googlered}{\scshape Interesting} are higher than those of the other factors. Most other scores fall somewhere in the middle, with factors like \textcolor{googlered}{\scshape Aesthetic Pleasure} having lower errors and factors like \textcolor{googlegreen}{\scshape Ease of Use} having higher errors.

The average accuracy of MLLMs over all factors ranges from $35\%$ to $38\%$, with \textbf{Claude having the highest accuracy of $38\%$. }The average $\pm 1$-accuracy is significantly higher, from $72\%$ to $77\%$. Again, the highest value is achieved by Claude. While the accuracy can vary a lot between different factors, we note that the $\pm 1$-accuracy is relatively stable. In terms of individual factors, Claude is either the best MLLM or performs similarly to it. The model is followed by Llama while GPT-4o performs slightly worse.

\subsubsection{Ranking Prediction}
\label{sec:ranking prediction}

Next, we analyze how well MLLMs predict the ranking of UIs by humans. All metrics are computed on average human and median MLLM scores, for each UI and factor, as in \cref{sec:individual scores}. The average scores are calculated by aggregating all evaluator scores by factor and UI, and taking the mean of those values.
For ranking scores, we consider three most popular metrics of correlation: Pearson correlation coefficient, Spearman's $\rho$, and Kendall's $\tau$. The \emph{Pearson correlation coefficient} is the covariance of MLLM and human scores divided by their standard deviations, over all UIs and for a given factor. The metric is in $[-1, 1]$, where $1$ is the maximum positive correlation, $-1$ is the maximum negative correlation, and $0$ means no correlations. Spearman's $\rho$ is the Pearson correlation coefficient applied to ranks of MLLM and human scores. Finally, Kendall's $\tau$ measures the pairwise agreement between scores ordered by MLLMs and humans, over all pairs of UIs and for a given factor. The metric is in $[-1, 1]$, where $1$ is the maximum positive correlation, $-1$ is the maximum negative correlation, and $0$ means no correlations. All metrics are reported in \cref{Table:FactorsScores}.

\cref{Table:FactorsScores} shows high correlation coefficients for multiple factors: \textcolor{googleblue}{\scshape Memorable}, \textcolor{googlered}{\scshape Aesthetic Pleasure}, and \textcolor{googlered}{\scshape Interesting}. On average, Llama has the highest correlation coefficients, followed by Claude and then GPT-4o. Some results are quite strong. For instance, the Kendall's $\tau$ of $0.4$, $0.6$, and $0.8$ means that the MLLM agrees with the human pairwise ranking $70\%$, $80\%$, and $90\%$ of the time, respectively.

Interestingly, high correlation metrics do not necessarily correlate with low prediction errors. As an example, factor \textcolor{googlered}{\scshape Interesting} has the highest MSE and MAE of all factors, and also low accuracy and $\pm 1$-accuracy. At the same time, it has the highest correlation metrics for all MLLMs. This reveals that although MLLMs cannot reliably predict the human score, they are effective at capturing relative preferences and thus useful for ranking UIs. This observation motivated us to take a deeper look at pairwise UI comparisons to examine whether these models are genuinely useful for ranking or preference-based tasks (\cref{sec:exp-pairwise-comparison}). The opposite can also be true. Factor \textcolor{googlered}{\scshape Comfort} has the lowest MSE and MAE of all factors, and also high $\pm 1$-accuracy. At the same time, all correlation metrics are close to zero. To conclude, both prediction errors and ranking metrics provide different views on predicted scores, and therefore we discuss both.

\subsection{Task 2: Pairwise Preference Prediction}
\label{sec:exp-pairwise-comparison}

In \cref{sec:exp-score-pred}, we evaluated the ability of MLLMs to predict human scores and their utility for estimating human preferences. In this section, we evaluate the ability of MLLMs to directly estimate human preferences. In \cref{sec:preference prediction}, we show that MLLMs can predict human preferences in some factors better than in others, as in \cref{Table:FactorsScores}. In \cref{sec:preference prediction hardness}, we show that human preferences are easier to predict if the compared UIs are judged as more different by humans.

\begin{table*}[h]
\centering
\small
\setlength{\tabcolsep}{1.00em}
\renewcommand{\arraystretch}{1.0}
\caption{Accuracy in predicting pairwise human preferences over UIs, for all factors and MLLM judges. The last column shows the overall accuracy, averaged over all factors. The highest accuracy, for both individual factors and overall, is reported in bold.}
\resizebox{0.9\textwidth}{!}{%
\begin{tabular}{lccccccccc|c}
\toprule
Model & \textcolor{googlegreen}{\bf Ease} & \textcolor{googlegreen}{\bf Clar} & \textcolor{googlegreen}{\bf Vis} & 
\textcolor{googleblue}{\bf Mem} & \textcolor{googleblue}{\bf Trust} & \textcolor{googleblue}{\bf Intui} & 
\textcolor{googlered}{\bf Aesth} & \textcolor{googlered}{\bf Intrst} & \textcolor{googlered}{\bf Comf} & \textbf{\sc \bfseries ALL} \\
\midrule
Claude-3.5 & 47.01 & \textbf{54.98} & \textbf{61.35} & 61.75 & \textbf{61.75} & 58.17 & \textbf{68.53} & \textbf{78.49} & 47.81 & \textbf{59.98} \\
GPT-4o & 50.99 & 54.55 & 56.92 & \textbf{64.82} & 58.89 & \textbf{59.68} & 64.03 & 75.10 & 51.38 & 59.60 \\
Llama-3.2 & \textbf{51.57} & 52.02 & 53.81 & 54.26 & 56.05 & 51.12 & 52.47 & 52.47 & \textbf{52.91} & 52.96 \\
\bottomrule
\end{tabular}%
}
\label{table:comparison_accuracy}
\vspace{-4pt}
\end{table*}

\begin{figure}[h]
\centering
\includegraphics[width=1.0\linewidth]{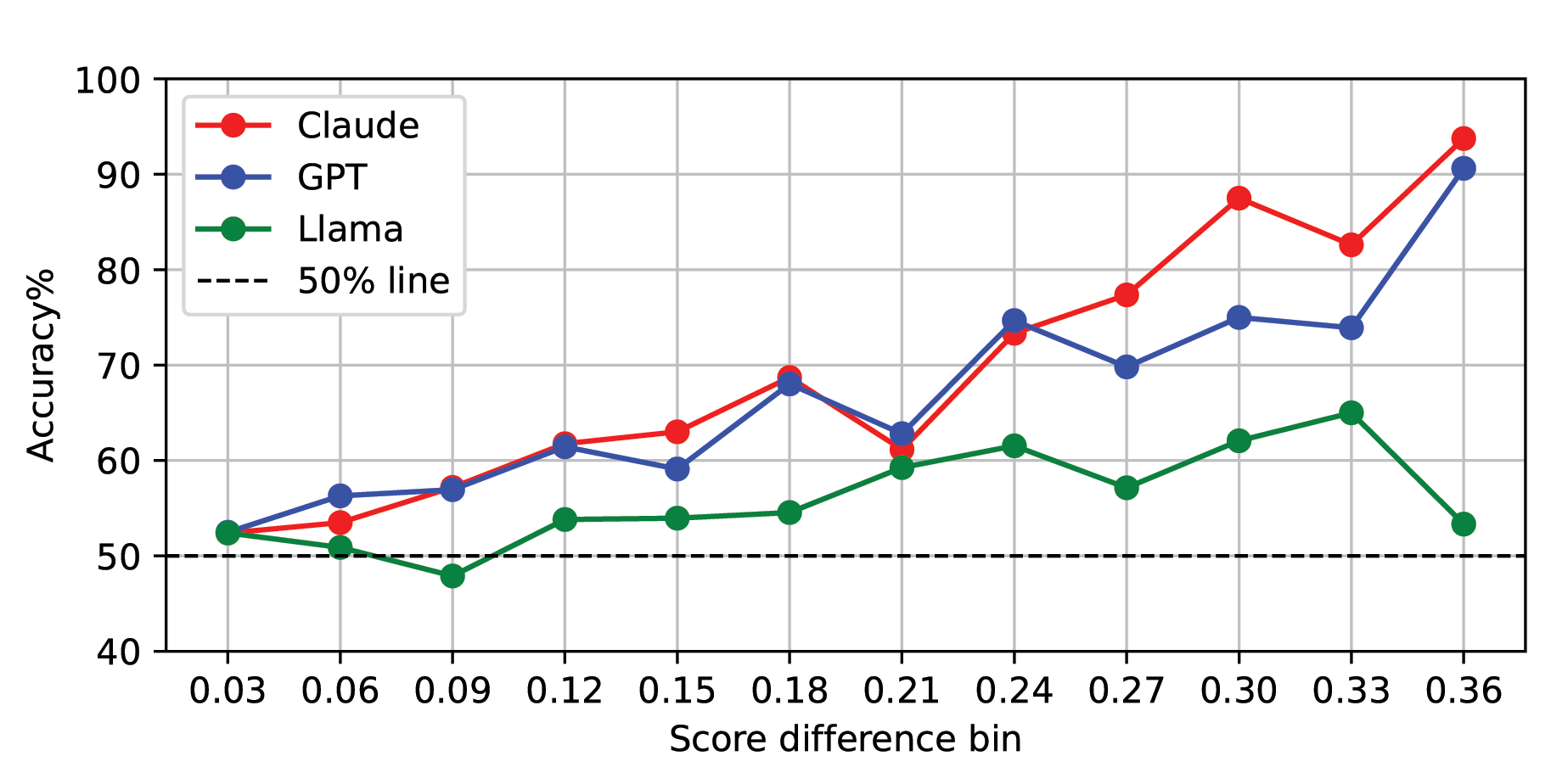}
\caption{Accuracy in predicting pairwise human preferences over UIs as a function of the absolute difference of their average human scores. A higher accuracy generally correlates with a larger difference in the human scores. The score-difference bin values were calculated as the difference between the average human ratings of the two UIs. This number is used to infer the difficulty of the question: smaller values indicate more difficult questions, while larger values imply easier ones.}
\label{fig:PairwiseAgreement}
\end{figure}

\subsubsection{Preference Prediction}
\label{sec:preference prediction}

This experiment is conducted as follows. First, we estimate the ground-truth human preference using the average human scores in \cref{sec:individual scores}. Specifically, for each pair of UIs for a given factor, we take the average human scores and say that UI $A$ is preferred to UI $B$ if the average score of the former is higher than that of the latter. Second, we obtain the preference of the MLLM judge. Specifically, for each pair of UIs, we run the judge with the prompt in \cref{fig:prompt-pairwise} and record its preference in each of the nine UI factors. Finally, we say that the judge agrees with the human preference over two UIs in a given factor if it prefers the same UI. The accuracy of the MLLM judge in a given factor is the percentage of UI pairs with agreement, and the overall accuracy is the average of these accuracies.

All accuracies are reported in \cref{table:comparison_accuracy}. They can vary drastically among the factors. GPT-4o and Claude both perform best when asked which UI is more \textcolor{googlered}{\scshape Interesting}, with accuracies around $75\%$ and $78\%$, respectively. This is similar to \cref{Table:FactorsScores}, where factor \textcolor{googlered}{\scshape Interesting} has the highest correlation coefficients and thus is highly predictable. All models perform poorly when judging \textcolor{googlegreen}{\scshape Ease of Use}, with the accuracies around $50\%$ and effectively indicating unpredictability. This is similar to \cref{Table:FactorsScores}, where factor \textcolor{googlegreen}{\scshape Ease of Use} has near-zero correlation coefficients. Both Claude and GPT-4o stand out in some areas while falling short in others. Llama is close to random across all factors. The overall accuracy reflects this, with both Claude and GPT-4o being close to $59\%$ and Llama being close to a random $51.79\%$.

\subsubsection{Hardness of Preference Prediction}
\label{sec:preference prediction hardness}

Although many of our UIs are professionally designed, humans can judge them very differently depending on the factor. This motivated us to study whether human preferences are easier to predict if the compared UIs are judged as more different by humans. Specifically, for all UI pairs and factors, we compute the absolute difference of their average human scores and bin them. For each bin, we report the accuracy of the MLLM judge to predict the human preference. 
We expect MLLMs to struggle more when predicting preferences for UI pairs that humans judged as more similar, but to perform more accurately when humans expressed a clear preference with larger rating differences.

Our results are reported in \cref{fig:PairwiseAgreement}. 
When the difference in human preferences between UI pairs is small, the accuracy of all MLLM judges is only slightly above $50\%$. This suggests that when humans expressed similar levels of preference for the two UIs, their choices were effectively unpredictable by MLLMs.
However, as the human preference score difference increases, suggesting that the UIs are more distinguishable for humans, the accuracy of the MLLM judges increases. 
GPT-4o and Claude have the largest increases, starting near $50\%$ and ending close to $90\%$ and $93\%$, respectively. Llama performs poorly, and its accuracy increases to only about $60\%$.

\section{Discussion}
\label{sec:discussion}

\subsection{MLLMs as Approximators}

Across all three models, the benchmark data illustrate that models can achieve reasonably high accuracy within $\pm1$ for absolute score tasks (\cref{table:comparison_accuracy}) and GPT and Claude perform reasonably well in pairwise tasks where the two UIs have a significant degree of difference (\cref{fig:PairwiseAgreement}). However, the same models perform less effectively in terms of exact accuracy (Claude 48\% and GPT 40\%) and when there is a small degree of difference in pairwise tasks. In fact, the models performed near random on pairwise tasks with small degrees of difference. Further, MAE scores ranged from 0.53 to 0.64, and MSE scores ranged from 0.44 to 0.66. While these scores are moderate, they are also significant enough to reflect that state of the art models cannot fully replace human evaluation. 

For this reason, it is recommended that MLLMs be seen as UI testing \textit{approximators}, not replacements. Essentially, given the results, MLLMs should be used to supplement existing research or be used in scenarios where research would not otherwise be used. The results of this study reflect that MLLMs struggled on fine-grained UI evaluation tasks. However, MLLMs have reasonable accuracy and reliability in generalized evaluations and when comparing UIs with larger disparities. Furthermore, results should be seen as approximations and are capable of giving designers general direction. However, they are not capable of acting as complete replacements for real human research.

\subsection{MLLMs' Role in the Design Process}
Given the moderate strength in human emulating UI tasks, MLLMs show decent promise in acting as a proxy for early user testing. Given that overall correlation coefficients across all models (\cref{table:comparison_accuracy}) were relatively decent, MLLMs show moderate promise for low-stakes user research. More specifically, designers should use them early in the design process in situations where they may be rapidly iterating through several different designs and not have the research resources to test and compare every single design. As seen in the results, MLLMs perform moderately well in these use cases as they can accurately score designs within $\pm1$ Likert scale point. Furthermore, as seen in the pairwise testing, they are also able to strongly predict human preferences when there is a large difference in human scores. For this reason, MLLMs can responsibly be used in these early design process situations in order to get a sense of direction on early UI iterations.

However, these results also suggest that MLLMs are not accurate enough to be used as a permanent or effective standout when exact results are needed. There was significant variation when looking at the results on the factor level, and it is clear that these models are not precise across most factors. For this reason, it is not recommended that these models be used for validation testing towards the end of the design process or high-stakes user research scenarios.

\subsection{Benchmark for Fine-Tuned Models}
Another future direction for this work would be to create a fine-tuned model with stronger factor sensitivity than the state-of-the-art models presented in this study. Considering that the state-of-the-art models mostly struggled with several traits like ease-of-use and even clarity, there is clearly a gap in how these models handle different traits. This gap suggests an opportunity to create a fine-tuned model based on the benchmark scores that will be better able to predict and replicate human scores. This benchmark can be utilized in tandem with reinforcement learning with human feedback (RLHF) on factor-specific data or even chain of thought prompting. Given this, the fine-tuned tool would be more useful in general, being more applicable and reliable in the design process. Furthermore, the data could be used to fine-tune prompts used. Doing so would require less lift than creating a fine-tuned model while also potentially making a significant difference in the evaluation quality. 

\subsection{Ethical Considerations}
While these results show encouraging reasons for using MLLMs as a UI judge, it should not be used to justify the complete replacement of human testing. In fact, it should do the opposite and reinforce the need for real human testing. The results presented expose key limitation that should give any design team pause when asking whether they should completely rely on MLLMs for user interface validation. 

Again, we emphasize that these results reflect that MLLMs should be used solely for supplemental use. It should be used to fill gaps where no user research exists, not to replace real human research and testing. MLLMs should be seen as a source of low-cost, rapid iterative evaluations when it is not feasible to perform extensive user research. This will enable designers to identify common trends and larger flaws earlier on in the design process.

Despite the seemingly negative ethical considerations, there is a strong reason for why this method should be adopted. There is significant potential for using this dataset to increase design democratization for junior designers with little to no access to high-end design research. Specifically, there is a large opportunity for designers to use MLLMs to "fine-tune" their designs, using MLLMs as a sanity check of their design intuitions and a learning opportunity.

\subsection{Future Work \& Limitations}

Building on our benchmark, we are extremely excited by the opportunities to expand AI’s capabilities as a judge for interface testing. Most notably, this benchmark can support fine-tuned models or alignment training—particularly reinforcement learning from human feedback (RLHF). Whether used alone or with similar datasets, it can help create models that better supplement user research. There is also the opportunity to expand the benchmark by adding more screens and tasks. For instance, this study focused on mobile and desktop UIs, but future work could include tablets, TVs, consoles, and more. A broader set of UIs would enable a more holistic benchmark. We also see potential to study more models and prompting methods—this work focused on GPT-4o, Claude 3.5 Sonnet, and Llama 3.2, but future studies could explore Gemini, Qwen, GPT-4o mini, and others using the same framework. Prompting strategies like chain-of-thought and few-shot learning also merit further exploration. Finally, because UI design is iterative, future work could compare versions of the same design, measuring how small changes—like color, text, or layout—affect human and MLLM judgments.

While our findings reveal MLLM's capability to emulate human UI perceptions to a moderate degree, several limitations of this study still persist. Firstly, the entirety of both the human and Firstly, the entirety of both the human and MLLMs tasks was performed on static user interfaces. Secondly, the screens used in the study were primarily professionally made within the context of Western design standards. We leave it to future extending our line of work to the diverse, dynamic interactions that usually occur in user research.

\section{Conclusion}
To evaluate the effectiveness of MLLMs as UI judges, this study facilitated the collection and testing of a set of user interfaces. 
The same UI studies were given to both human and MLLM evaluators, and the results were compared to analyze to what degree MLLM evaluators can emulate human results. This study was conducted using two unique tasks: 1) an absolute scoring task using a 1-7 Likert scale and 2) a pairwise evaluation. For task 1, the findings reveal that when asked to evaluate screens at an overall level, MLLMs do moderately well. However, when asked to judge them based on specific factors or criteria, they tend to struggle. For task 2, the MLLMs again performed reasonably on the pairwise comparisons. Naturally, MLLM evaluators performed better on pairwise questions that compared two screens that have a higher degree of difference. Lastly, we accumulated the results from these tests into a benchmark to be used in future studies.
Future work offers the opportunity to use the benchmark to create a fine-tuned model that would perform better at UI judgment tasks compared to state-of-the-art models. For this reason, we recommend that MLLMs are used to supplement early, low stakes user research -- not replace it. Finally, the benchmark provided in this paper is capable of laying the groundwork for future studies that can better explore MLLM as a UI judge.

\bibliographystyle{ACM-Reference-Format}
\bibliography{main}


\begin{thebibliography}{63}

%%% the Web address.

\ifx \showCODEN    \undefined \def \showCODEN     #1{\unskip}     \fi
\ifx \showDOI      \undefined \def \showDOI       #1{#1}\fi
\ifx \showISBNx    \undefined \def \showISBNx     #1{\unskip}     \fi
\ifx \showISBNxiii \undefined \def \showISBNxiii  #1{\unskip}     \fi
\ifx \showISSN     \undefined \def \showISSN      #1{\unskip}     \fi
\ifx \showLCCN     \undefined \def \showLCCN      #1{\unskip}     \fi
\ifx \shownote     \undefined \def \shownote      #1{#1}          \fi
\ifx \showarticletitle \undefined \def \showarticletitle #1{#1}   \fi
\ifx \showURL      \undefined \def \showURL       {\relax}        \fi
% invisible to TeX
\providecommand\bibfield[2]{#2}
\providecommand\bibinfo[2]{#2}
\providecommand\natexlab[1]{#1}
\providecommand\showeprint[2][]{arXiv:#2}

\bibitem[Achiam et~al\mbox{.}(2023)]%
        {achiam2023gpt}
\bibfield{author}{\bibinfo{person}{Josh Achiam}, \bibinfo{person}{Steven Adler}, \bibinfo{person}{Sandhini Agarwal}, \bibinfo{person}{Lama Ahmad}, \bibinfo{person}{Ilge Akkaya}, \bibinfo{person}{Florencia~Leoni Aleman}, \bibinfo{person}{Diogo Almeida}, \bibinfo{person}{Janko Altenschmidt}, \bibinfo{person}{Sam Altman}, \bibinfo{person}{Shyamal Anadkat}, {et~al\mbox{.}}} \bibinfo{year}{2023}\natexlab{}.
\newblock \showarticletitle{Gpt-4 technical report}.
\newblock \bibinfo{journal}{\emph{arXiv preprint arXiv:2303.08774}} (\bibinfo{year}{2023}).
\newblock


\bibitem[B{\ae}rentsen(2000)]%
        {baerentsen2000intuitive}
\bibfield{author}{\bibinfo{person}{Klaus~B B{\ae}rentsen}.} \bibinfo{year}{2000}\natexlab{}.
\newblock \showarticletitle{Intuitive user interfaces}.
\newblock \bibinfo{journal}{\emph{Scandinavian Journal of Information Systems}} \bibinfo{volume}{12}, \bibinfo{number}{1} (\bibinfo{year}{2000}), \bibinfo{pages}{4}.
\newblock


\bibitem[Bailey et~al\mbox{.}(1988)]%
        {bailey1988effects}
\bibfield{author}{\bibinfo{person}{Wayne~A Bailey}, \bibinfo{person}{Stephen~T Knox}, {and} \bibinfo{person}{Eugene~F Lynch}.} \bibinfo{year}{1988}\natexlab{}.
\newblock \showarticletitle{Effects of interface design upon user productivity}. In \bibinfo{booktitle}{\emph{Proceedings of the SIGCHI conference on Human factors in computing systems}}. \bibinfo{pages}{207--212}.
\newblock


\bibitem[Baxter et~al\mbox{.}(2015)]%
        {baxter2015understanding}
\bibfield{author}{\bibinfo{person}{Kathy Baxter}, \bibinfo{person}{Catherine Courage}, {and} \bibinfo{person}{Kelly Caine}.} \bibinfo{year}{2015}\natexlab{}.
\newblock \bibinfo{booktitle}{\emph{Understanding your users: a practical guide to user research methods}}.
\newblock \bibinfo{publisher}{Morgan Kaufmann}.
\newblock


\bibitem[Bert{\~a}o and Joo(2021)]%
        {bertao2021artificial}
\bibfield{author}{\bibinfo{person}{Renato~Antonio Bert{\~a}o} {and} \bibinfo{person}{Jaewoo Joo}.} \bibinfo{year}{2021}\natexlab{}.
\newblock \showarticletitle{Artificial intelligence in UX/UI design: a survey on current adoption and [future] practices}.
\newblock \bibinfo{journal}{\emph{Safe Harbors for Design Research}} (\bibinfo{year}{2021}), \bibinfo{pages}{1--10}.
\newblock


\bibitem[Bhandari et~al\mbox{.}(2017)]%
        {bhandari2017effects}
\bibfield{author}{\bibinfo{person}{Upasna Bhandari}, \bibinfo{person}{Tillmann Neben}, \bibinfo{person}{Klarissa Chang}, {and} \bibinfo{person}{Wen~Yong Chua}.} \bibinfo{year}{2017}\natexlab{}.
\newblock \showarticletitle{Effects of interface design factors on affective responses and quality evaluations in mobile applications}.
\newblock \bibinfo{journal}{\emph{Computers in Human Behavior}}  \bibinfo{volume}{72} (\bibinfo{year}{2017}), \bibinfo{pages}{525--534}.
\newblock


\bibitem[Blackler et~al\mbox{.}(2005)]%
        {blackler2005intuitive}
\bibfield{author}{\bibinfo{person}{Alethea Blackler}, \bibinfo{person}{Vesna Popovic}, {and} \bibinfo{person}{Douglas Mahar}.} \bibinfo{year}{2005}\natexlab{}.
\newblock \showarticletitle{Intuitive interaction applied to interface design}. In \bibinfo{booktitle}{\emph{New Design Paradigms: Proceedings of International Design Congress (IDC) 2005}}. International Design Congress, \bibinfo{pages}{1--10}.
\newblock


\bibitem[Bollini(2017)]%
        {bollini2017beautiful}
\bibfield{author}{\bibinfo{person}{Letizia Bollini}.} \bibinfo{year}{2017}\natexlab{}.
\newblock \showarticletitle{Beautiful interfaces. From user experience to user interface design}.
\newblock \bibinfo{journal}{\emph{The Design Journal}} \bibinfo{volume}{20}, \bibinfo{number}{sup1} (\bibinfo{year}{2017}), \bibinfo{pages}{S89--S101}.
\newblock


\bibitem[Borkin et~al\mbox{.}(2013)]%
        {6634103}
\bibfield{author}{\bibinfo{person}{Michelle~A. Borkin}, \bibinfo{person}{Azalea~A. Vo}, \bibinfo{person}{Zoya Bylinskii}, \bibinfo{person}{Phillip Isola}, \bibinfo{person}{Shashank Sunkavalli}, \bibinfo{person}{Aude Oliva}, {and} \bibinfo{person}{Hanspeter Pfister}.} \bibinfo{year}{2013}\natexlab{}.
\newblock \showarticletitle{What Makes a Visualization Memorable?}
\newblock \bibinfo{journal}{\emph{IEEE Transactions on Visualization and Computer Graphics}} \bibinfo{volume}{19}, \bibinfo{number}{12} (\bibinfo{year}{2013}), \bibinfo{pages}{2306--2315}.
\newblock
\urldef\tempurl%
\url{https://doi.org/10.1109/TVCG.2013.234}
\showDOI{\tempurl}


\bibitem[Cao et~al\mbox{.}(2025)]%
        {cao2025survey}
\bibfield{author}{\bibinfo{person}{Renjie Cao}, \bibinfo{person}{Mei Zhang}, {and} \bibinfo{person}{Yu Huang}.} \bibinfo{year}{2025}\natexlab{}.
\newblock \showarticletitle{Learning to Simulate Survey Distributions with Fine-Tuned LLMs}. In \bibinfo{booktitle}{\emph{Proceedings of the 2025 Conference of the North American Chapter of the Association for Computational Linguistics (NAACL)}}.
\newblock


\bibitem[Darejeh et~al\mbox{.}(2024)]%
        {darejeh2024critical}
\bibfield{author}{\bibinfo{person}{Ali Darejeh}, \bibinfo{person}{Nadine Marcusa}, \bibinfo{person}{Gelareh Mohammadi}, {and} \bibinfo{person}{John Sweller}.} \bibinfo{year}{2024}\natexlab{}.
\newblock \showarticletitle{A critical analysis of cognitive load measurement methods for evaluating the usability of different types of interfaces: guidelines and framework for Human-Computer Interaction}.
\newblock \bibinfo{journal}{\emph{arXiv preprint arXiv:2402.11820}} (\bibinfo{year}{2024}).
\newblock


\bibitem[Das et~al\mbox{.}(2025)]%
        {das2025leveraging}
\bibfield{author}{\bibinfo{person}{Amit~Kumar Das}, \bibinfo{person}{Cindy~Xiong Bearfield}, {and} \bibinfo{person}{Klaus Mueller}.} \bibinfo{year}{2025}\natexlab{}.
\newblock \showarticletitle{Leveraging Large Language Models for Personalized Public Messaging}. In \bibinfo{booktitle}{\emph{Proceedings of the Extended Abstracts of the CHI Conference on Human Factors in Computing Systems}}. \bibinfo{pages}{1--7}.
\newblock


\bibitem[Dave et~al\mbox{.}(2023)]%
        {dave2023understanding}
\bibfield{author}{\bibinfo{person}{Anjali Dave}, \bibinfo{person}{Ankur Saxena}, {and} \bibinfo{person}{Avdhesh Jha}.} \bibinfo{year}{2023}\natexlab{}.
\newblock \showarticletitle{Understanding User comfort and Expectations in AI-based Systems}.
\newblock  (\bibinfo{year}{2023}).
\newblock


\bibitem[Duan et~al\mbox{.}(2024a)]%
        {duan2024uicrit}
\bibfield{author}{\bibinfo{person}{Peitong Duan}, \bibinfo{person}{Chin-Yi Cheng}, \bibinfo{person}{Gang Li}, \bibinfo{person}{Bjoern Hartmann}, {and} \bibinfo{person}{Yang Li}.} \bibinfo{year}{2024}\natexlab{a}.
\newblock \showarticletitle{UICrit: Enhancing Automated Design Evaluation with a UI Critique Dataset}. In \bibinfo{booktitle}{\emph{Proceedings of the 37th Annual ACM Symposium on User Interface Software and Technology}}. \bibinfo{pages}{1--17}.
\newblock


\bibitem[Duan et~al\mbox{.}(2024b)]%
        {duan2024generating}
\bibfield{author}{\bibinfo{person}{Peitong Duan}, \bibinfo{person}{Jeremy Warner}, \bibinfo{person}{Yang Li}, {and} \bibinfo{person}{Bjoern Hartmann}.} \bibinfo{year}{2024}\natexlab{b}.
\newblock \showarticletitle{Generating automatic feedback on ui mockups with large language models}. In \bibinfo{booktitle}{\emph{Proceedings of the 2024 CHI Conference on Human Factors in Computing Systems}}. \bibinfo{pages}{1--20}.
\newblock


\bibitem[Gu et~al\mbox{.}(2024)]%
        {gu2024survey}
\bibfield{author}{\bibinfo{person}{Jiawei Gu}, \bibinfo{person}{Xuhui Jiang}, \bibinfo{person}{Zhichao Shi}, \bibinfo{person}{Hexiang Tan}, \bibinfo{person}{Xuehao Zhai}, \bibinfo{person}{Chengjin Xu}, \bibinfo{person}{Wei Li}, \bibinfo{person}{Yinghan Shen}, \bibinfo{person}{Shengjie Ma}, \bibinfo{person}{Honghao Liu}, {et~al\mbox{.}}} \bibinfo{year}{2024}\natexlab{}.
\newblock \showarticletitle{A survey on llm-as-a-judge}.
\newblock \bibinfo{journal}{\emph{arXiv preprint arXiv:2411.15594}} (\bibinfo{year}{2024}).
\newblock


\bibitem[Guo et~al\mbox{.}(2025)]%
        {guo2025deepseek}
\bibfield{author}{\bibinfo{person}{Daya Guo}, \bibinfo{person}{Dejian Yang}, \bibinfo{person}{Haowei Zhang}, \bibinfo{person}{Junxiao Song}, \bibinfo{person}{Ruoyu Zhang}, \bibinfo{person}{Runxin Xu}, \bibinfo{person}{Qihao Zhu}, \bibinfo{person}{Shirong Ma}, \bibinfo{person}{Peiyi Wang}, \bibinfo{person}{Xiao Bi}, {et~al\mbox{.}}} \bibinfo{year}{2025}\natexlab{}.
\newblock \showarticletitle{Deepseek-r1: Incentivizing reasoning capability in llms via reinforcement learning}.
\newblock \bibinfo{journal}{\emph{arXiv preprint arXiv:2501.12948}} (\bibinfo{year}{2025}).
\newblock


\bibitem[H\"{a}m\"{a}l\"{a}inen et~al\mbox{.}(2023)]%
        {10.1145/3544548.3580688}
\bibfield{author}{\bibinfo{person}{Perttu H\"{a}m\"{a}l\"{a}inen}, \bibinfo{person}{Mikke Tavast}, {and} \bibinfo{person}{Anton Kunnari}.} \bibinfo{year}{2023}\natexlab{}.
\newblock \showarticletitle{Evaluating Large Language Models in Generating Synthetic HCI Research Data: a Case Study}. In \bibinfo{booktitle}{\emph{Proceedings of the 2023 CHI Conference on Human Factors in Computing Systems}} (Hamburg, Germany) \emph{(\bibinfo{series}{CHI '23})}. \bibinfo{publisher}{Association for Computing Machinery}, \bibinfo{address}{New York, NY, USA}, Article \bibinfo{articleno}{433}, \bibinfo{numpages}{19}~pages.
\newblock
\showISBNx{9781450394215}
\urldef\tempurl%
\url{https://doi.org/10.1145/3544548.3580688}
\showDOI{\tempurl}


\bibitem[Hartmann et~al\mbox{.}(2008)]%
        {hartmann2008towards}
\bibfield{author}{\bibinfo{person}{Jan Hartmann}, \bibinfo{person}{Alistair Sutcliffe}, {and} \bibinfo{person}{Antonella~De Angeli}.} \bibinfo{year}{2008}\natexlab{}.
\newblock \showarticletitle{Towards a theory of user judgment of aesthetics and user interface quality}.
\newblock \bibinfo{journal}{\emph{ACM Transactions on Computer-Human Interaction (TOCHI)}} \bibinfo{volume}{15}, \bibinfo{number}{4} (\bibinfo{year}{2008}), \bibinfo{pages}{1--30}.
\newblock


\bibitem[Hinton(2012)]%
        {hinton2012using}
\bibfield{author}{\bibinfo{person}{Stephen Hinton}.} \bibinfo{year}{2012}\natexlab{}.
\newblock \showarticletitle{Using Recursive Distance Vector Methodology to Review Remote Desktop Solutions in the Small Business Consulting Environment}.
\newblock \bibinfo{journal}{\emph{Issues in Information Systems}} \bibinfo{volume}{13}, \bibinfo{number}{1} (\bibinfo{year}{2012}), \bibinfo{pages}{94--104}.
\newblock


\bibitem[Jansen et~al\mbox{.}(2023)]%
        {DBLP:journals/nlpj/JansenJS23}
\bibfield{author}{\bibinfo{person}{Bernard~J. Jansen}, \bibinfo{person}{Soon-Gyo Jung}, {and} \bibinfo{person}{Joni Salminen}.} \bibinfo{year}{2023}\natexlab{}.
\newblock \showarticletitle{Employing large language models in survey research}.
\newblock \bibinfo{journal}{\emph{Nat. Lang. Process. J.}}  \bibinfo{volume}{4} (\bibinfo{year}{2023}), \bibinfo{pages}{100020}.
\newblock
\urldef\tempurl%
\url{https://doi.org/10.1016/j.nlp.2023.100020}
\showURL{%
\tempurl}


\bibitem[Kaplan(2020)]%
        {kaplan2020}
\bibfield{author}{\bibinfo{person}{Kate Kaplan}.} \bibinfo{year}{2020}\natexlab{}.
\newblock \bibinfo{title}{Typical Designer–to–Developer and Researcher–to–Designer Ratios}.
\newblock
\newblock
\urldef\tempurl%
\url{https://www.nngroup.com/articles/ux-developer-ratio/}
\showURL{%
\tempurl}


\bibitem[Kim et~al\mbox{.}(2024)]%
        {kim2024prometheus}
\bibfield{author}{\bibinfo{person}{Seungone Kim}, \bibinfo{person}{Juyoung Suk}, \bibinfo{person}{Shayne Longpre}, \bibinfo{person}{Bill~Yuchen Lin}, \bibinfo{person}{Jamin Shin}, \bibinfo{person}{Sean Welleck}, \bibinfo{person}{Graham Neubig}, \bibinfo{person}{Moontae Lee}, \bibinfo{person}{Kyungjae Lee}, {and} \bibinfo{person}{Minjoon Seo}.} \bibinfo{year}{2024}\natexlab{}.
\newblock \showarticletitle{Prometheus 2: An open source language model specialized in evaluating other language models}.
\newblock \bibinfo{journal}{\emph{arXiv preprint arXiv:2405.01535}} (\bibinfo{year}{2024}).
\newblock


\bibitem[Kim et~al\mbox{.}(2025)]%
        {kim2025chart}
\bibfield{author}{\bibinfo{person}{Seon~Gyeom Kim}, \bibinfo{person}{Jae~Young Choi}, \bibinfo{person}{Ryan Rossi}, \bibinfo{person}{Eunyee Koh}, {and} \bibinfo{person}{Tak~Yeon Lee}.} \bibinfo{year}{2025}\natexlab{}.
\newblock \showarticletitle{Chart-to-Experience: Benchmarking Multimodal LLMs for Predicting Experiential Impact of Charts}. In \bibinfo{booktitle}{\emph{2025 IEEE 18th Pacific Visualization Conference (PacificVis)}}. IEEE, \bibinfo{pages}{340--345}.
\newblock


\bibitem[Kuang et~al\mbox{.}(2024)]%
        {10.1145/3613904.3642168}
\bibfield{author}{\bibinfo{person}{Emily Kuang}, \bibinfo{person}{Minghao Li}, \bibinfo{person}{Mingming Fan}, {and} \bibinfo{person}{Kristen Shinohara}.} \bibinfo{year}{2024}\natexlab{}.
\newblock \showarticletitle{Enhancing UX Evaluation Through Collaboration with Conversational AI Assistants: Effects of Proactive Dialogue and Timing}. In \bibinfo{booktitle}{\emph{Proceedings of the 2024 CHI Conference on Human Factors in Computing Systems}} (Honolulu, HI, USA) \emph{(\bibinfo{series}{CHI '24})}. \bibinfo{publisher}{Association for Computing Machinery}, \bibinfo{address}{New York, NY, USA}, Article \bibinfo{articleno}{3}, \bibinfo{numpages}{16}~pages.
\newblock
\showISBNx{9798400703300}
\urldef\tempurl%
\url{https://doi.org/10.1145/3613904.3642168}
\showDOI{\tempurl}


\bibitem[Kumar et~al\mbox{.}(2004)]%
        {kumar2004user}
\bibfield{author}{\bibinfo{person}{Ram~L Kumar}, \bibinfo{person}{Michael~Alan Smith}, {and} \bibinfo{person}{Snehamay Bannerjee}.} \bibinfo{year}{2004}\natexlab{}.
\newblock \showarticletitle{User interface features influencing overall ease of use and personalization}.
\newblock \bibinfo{journal}{\emph{Information \& Management}} \bibinfo{volume}{41}, \bibinfo{number}{3} (\bibinfo{year}{2004}), \bibinfo{pages}{289--302}.
\newblock


\bibitem[Lee et~al\mbox{.}(2024)]%
        {lee2024prometheus}
\bibfield{author}{\bibinfo{person}{Seongyun Lee}, \bibinfo{person}{Seungone Kim}, \bibinfo{person}{Sue Park}, \bibinfo{person}{Geewook Kim}, {and} \bibinfo{person}{Minjoon Seo}.} \bibinfo{year}{2024}\natexlab{}.
\newblock \showarticletitle{Prometheus-vision: Vision-language model as a judge for fine-grained evaluation}. In \bibinfo{booktitle}{\emph{Findings of the association for computational linguistics ACL 2024}}. \bibinfo{pages}{11286--11315}.
\newblock


\bibitem[Li et~al\mbox{.}(2023)]%
        {li2023generative}
\bibfield{author}{\bibinfo{person}{Junlong Li}, \bibinfo{person}{Shichao Sun}, \bibinfo{person}{Weizhe Yuan}, \bibinfo{person}{Run-Ze Fan}, \bibinfo{person}{Hai Zhao}, {and} \bibinfo{person}{Pengfei Liu}.} \bibinfo{year}{2023}\natexlab{}.
\newblock \showarticletitle{Generative judge for evaluating alignment}.
\newblock \bibinfo{journal}{\emph{arXiv preprint arXiv:2310.05470}} (\bibinfo{year}{2023}).
\newblock


\bibitem[Li et~al\mbox{.}(2024)]%
        {li2024frontiers}
\bibfield{author}{\bibinfo{person}{Peiyao Li}, \bibinfo{person}{Noah Castelo}, \bibinfo{person}{Zsolt Katona}, {and} \bibinfo{person}{Miklos Sarvary}.} \bibinfo{year}{2024}\natexlab{}.
\newblock \showarticletitle{Frontiers: Determining the validity of large language models for automated perceptual analysis}.
\newblock \bibinfo{journal}{\emph{Marketing Science}} \bibinfo{volume}{43}, \bibinfo{number}{2} (\bibinfo{year}{2024}), \bibinfo{pages}{254--266}.
\newblock


\bibitem[Liu et~al\mbox{.}(2024)]%
        {liu2024aligning}
\bibfield{author}{\bibinfo{person}{Yinhong Liu}, \bibinfo{person}{Han Zhou}, \bibinfo{person}{Zhijiang Guo}, \bibinfo{person}{Ehsan Shareghi}, \bibinfo{person}{Ivan Vuli{\'c}}, \bibinfo{person}{Anna Korhonen}, {and} \bibinfo{person}{Nigel Collier}.} \bibinfo{year}{2024}\natexlab{}.
\newblock \showarticletitle{Aligning with human judgement: The role of pairwise preference in large language model evaluators}.
\newblock \bibinfo{journal}{\emph{arXiv preprint arXiv:2403.16950}} (\bibinfo{year}{2024}).
\newblock


\bibitem[Luera et~al\mbox{.}(2025)]%
        {10.1145/3701716.3715452}
\bibfield{author}{\bibinfo{person}{Reuben Luera}, \bibinfo{person}{Ryan Rossi}, \bibinfo{person}{Franck Dernoncourt}, \bibinfo{person}{Alexa Siu}, \bibinfo{person}{Sungchul Kim}, \bibinfo{person}{Tong Yu}, \bibinfo{person}{Ruiyi Zhang}, \bibinfo{person}{Xiang Chen}, \bibinfo{person}{Nedim Lipka}, \bibinfo{person}{Zhehao Zhang}, \bibinfo{person}{Seon Gyeom~Kim}, {and} \bibinfo{person}{Tak Yeon~Lee}.} \bibinfo{year}{2025}\natexlab{}.
\newblock \showarticletitle{Personalizing Data Delivery: Investigating User Characteristics and Enhancing LLM Predictions}. In \bibinfo{booktitle}{\emph{Companion Proceedings of the ACM on Web Conference 2025}} (Sydney NSW, Australia) \emph{(\bibinfo{series}{WWW '25})}. \bibinfo{publisher}{Association for Computing Machinery}, \bibinfo{address}{New York, NY, USA}, \bibinfo{pages}{1167–1171}.
\newblock
\showISBNx{9798400713316}
\urldef\tempurl%
\url{https://doi.org/10.1145/3701716.3715452}
\showDOI{\tempurl}


\bibitem[Luera et~al\mbox{.}(2024)]%
        {luera2024survey}
\bibfield{author}{\bibinfo{person}{Reuben Luera}, \bibinfo{person}{Ryan~A Rossi}, \bibinfo{person}{Alexa Siu}, \bibinfo{person}{Franck Dernoncourt}, \bibinfo{person}{Tong Yu}, \bibinfo{person}{Sungchul Kim}, \bibinfo{person}{Ruiyi Zhang}, \bibinfo{person}{Xiang Chen}, \bibinfo{person}{Hanieh Salehy}, \bibinfo{person}{Jian Zhao}, {et~al\mbox{.}}} \bibinfo{year}{2024}\natexlab{}.
\newblock \showarticletitle{Survey of User Interface Design and Interaction Techniques in Generative AI Applications}.
\newblock \bibinfo{journal}{\emph{arXiv preprint arXiv:2410.22370}} (\bibinfo{year}{2024}).
\newblock


\bibitem[Macaranas et~al\mbox{.}(2015)]%
        {macaranas2015intuitive}
\bibfield{author}{\bibinfo{person}{Anna Macaranas}, \bibinfo{person}{Alissa~N Antle}, {and} \bibinfo{person}{Bernhard~E Riecke}.} \bibinfo{year}{2015}\natexlab{}.
\newblock \showarticletitle{What is intuitive interaction? Balancing users’ performance and satisfaction with natural user interfaces}.
\newblock \bibinfo{journal}{\emph{Interacting with Computers}} \bibinfo{volume}{27}, \bibinfo{number}{3} (\bibinfo{year}{2015}), \bibinfo{pages}{357--370}.
\newblock


\bibitem[Medlock(2018)]%
        {medlock2018rapid}
\bibfield{author}{\bibinfo{person}{Michael~C Medlock}.} \bibinfo{year}{2018}\natexlab{}.
\newblock \showarticletitle{The rapid iterative test and evaluation method (RITE)}.
\newblock \bibinfo{journal}{\emph{Games User Research}} (\bibinfo{year}{2018}), \bibinfo{pages}{203--215}.
\newblock


\bibitem[Muzaffar et~al\mbox{.}(2011)]%
        {muzaffar2011usability}
\bibfield{author}{\bibinfo{person}{Abdul~Wahab Muzaffar}, \bibinfo{person}{Farooque Azam}, \bibinfo{person}{Hina Anwar}, {and} \bibinfo{person}{Ali~Saeed Khan}.} \bibinfo{year}{2011}\natexlab{}.
\newblock \showarticletitle{Usability aspects in pervasive computing: Needs and challenges}.
\newblock \bibinfo{journal}{\emph{International Journal of Computer Applications}} \bibinfo{volume}{32}, \bibinfo{number}{10} (\bibinfo{year}{2011}).
\newblock


\bibitem[Nielsen(2005)]%
        {nielsen2005ten}
\bibfield{author}{\bibinfo{person}{Jakob Nielsen}.} \bibinfo{year}{2005}\natexlab{}.
\newblock \showarticletitle{Ten usability heuristics}.
\newblock  (\bibinfo{year}{2005}).
\newblock


\bibitem[Ouyang et~al\mbox{.}(2022)]%
        {ouyang2022training}
\bibfield{author}{\bibinfo{person}{Long Ouyang}, \bibinfo{person}{Jeffrey Wu}, \bibinfo{person}{Xu Jiang}, \bibinfo{person}{Diogo Almeida}, \bibinfo{person}{Carroll Wainwright}, \bibinfo{person}{Pamela Mishkin}, \bibinfo{person}{Chong Zhang}, \bibinfo{person}{Sandhini Agarwal}, \bibinfo{person}{Katarina Slama}, \bibinfo{person}{Alex Ray}, {et~al\mbox{.}}} \bibinfo{year}{2022}\natexlab{}.
\newblock \showarticletitle{Training language models to follow instructions with human feedback}.
\newblock \bibinfo{journal}{\emph{Advances in neural information processing systems}}  \bibinfo{volume}{35} (\bibinfo{year}{2022}), \bibinfo{pages}{27730--27744}.
\newblock


\bibitem[Oviatt(2006)]%
        {oviatt2006human}
\bibfield{author}{\bibinfo{person}{Sharon Oviatt}.} \bibinfo{year}{2006}\natexlab{}.
\newblock \showarticletitle{Human-centered design meets cognitive load theory: designing interfaces that help people think}. In \bibinfo{booktitle}{\emph{Proceedings of the 14th ACM international conference on Multimedia}}. \bibinfo{pages}{871--880}.
\newblock


\bibitem[Panickssery et~al\mbox{.}(2024)]%
        {panickssery2024llmevaluatorsrecognizefavor}
\bibfield{author}{\bibinfo{person}{Arjun Panickssery}, \bibinfo{person}{Samuel~R. Bowman}, {and} \bibinfo{person}{Shi Feng}.} \bibinfo{year}{2024}\natexlab{}.
\newblock \bibinfo{title}{LLM Evaluators Recognize and Favor Their Own Generations}.
\newblock
\newblock
\showeprint[arxiv]{2404.13076}~[cs.CL]
\urldef\tempurl%
\url{https://arxiv.org/abs/2404.13076}
\showURL{%
\tempurl}


\bibitem[Qin et~al\mbox{.}(2024)]%
        {qin2024largelanguagemodelseffective}
\bibfield{author}{\bibinfo{person}{Zhen Qin}, \bibinfo{person}{Rolf Jagerman}, \bibinfo{person}{Kai Hui}, \bibinfo{person}{Honglei Zhuang}, \bibinfo{person}{Junru Wu}, \bibinfo{person}{Le Yan}, \bibinfo{person}{Jiaming Shen}, \bibinfo{person}{Tianqi Liu}, \bibinfo{person}{Jialu Liu}, \bibinfo{person}{Donald Metzler}, \bibinfo{person}{Xuanhui Wang}, {and} \bibinfo{person}{Michael Bendersky}.} \bibinfo{year}{2024}\natexlab{}.
\newblock \bibinfo{title}{Large Language Models are Effective Text Rankers with Pairwise Ranking Prompting}.
\newblock
\newblock
\showeprint[arxiv]{2306.17563}~[cs.IR]
\urldef\tempurl%
\url{https://arxiv.org/abs/2306.17563}
\showURL{%
\tempurl}


\bibitem[Ramamurthy et~al\mbox{.}(2022)]%
        {ramamurthy2022reinforcement}
\bibfield{author}{\bibinfo{person}{Rajkumar Ramamurthy}, \bibinfo{person}{Prithviraj Ammanabrolu}, \bibinfo{person}{Kiant{\'e} Brantley}, \bibinfo{person}{Jack Hessel}, \bibinfo{person}{Rafet Sifa}, \bibinfo{person}{Christian Bauckhage}, \bibinfo{person}{Hannaneh Hajishirzi}, {and} \bibinfo{person}{Yejin Choi}.} \bibinfo{year}{2022}\natexlab{}.
\newblock \showarticletitle{Is reinforcement learning (not) for natural language processing: Benchmarks, baselines, and building blocks for natural language policy optimization}.
\newblock \bibinfo{journal}{\emph{arXiv preprint arXiv:2210.01241}} (\bibinfo{year}{2022}).
\newblock


\bibitem[Reinecke et~al\mbox{.}(2013)]%
        {reinecke2013predicting}
\bibfield{author}{\bibinfo{person}{Katharina Reinecke}, \bibinfo{person}{Tom Yeh}, \bibinfo{person}{Luke Miratrix}, \bibinfo{person}{Rahmatri Mardiko}, \bibinfo{person}{Yuechen Zhao}, \bibinfo{person}{Jenny Liu}, {and} \bibinfo{person}{Krzysztof~Z Gajos}.} \bibinfo{year}{2013}\natexlab{}.
\newblock \showarticletitle{Predicting users' first impressions of website aesthetics with a quantification of perceived visual complexity and colorfulness}. In \bibinfo{booktitle}{\emph{Proceedings of the SIGCHI conference on human factors in computing systems}}. \bibinfo{pages}{2049--2058}.
\newblock


\bibitem[Rosala and Moran(2024)]%
        {rosala2024synthetic}
\bibfield{author}{\bibinfo{person}{Tina Rosala} {and} \bibinfo{person}{Kate Moran}.} \bibinfo{year}{2024}\natexlab{}.
\newblock \showarticletitle{Synthetic Users and AI Personas in UX Research: A Cautionary Note}.
\newblock \bibinfo{journal}{\emph{Nielsen Norman Group Reports}} (\bibinfo{year}{2024}).
\newblock
\urldef\tempurl%
\url{https://www.nngroup.com/articles/synthetic-users-ai/}
\showURL{%
\tempurl}


\bibitem[Schoop et~al\mbox{.}(2022)]%
        {Schoop_2022}
\bibfield{author}{\bibinfo{person}{Eldon Schoop}, \bibinfo{person}{Xin Zhou}, \bibinfo{person}{Gang Li}, \bibinfo{person}{Zhourong Chen}, \bibinfo{person}{Bjoern Hartmann}, {and} \bibinfo{person}{Yang Li}.} \bibinfo{year}{2022}\natexlab{}.
\newblock \showarticletitle{Predicting and Explaining Mobile UI Tappability with Vision Modeling and Saliency Analysis}. In \bibinfo{booktitle}{\emph{CHI Conference on Human Factors in Computing Systems}} \emph{(\bibinfo{series}{CHI ’22})}. \bibinfo{publisher}{ACM}, \bibinfo{pages}{1–21}.
\newblock
\urldef\tempurl%
\url{https://doi.org/10.1145/3491102.3517497}
\showDOI{\tempurl}


\bibitem[Seeley(2012)]%
        {seeley2012hearing}
\bibfield{author}{\bibinfo{person}{William~P Seeley}.} \bibinfo{year}{2012}\natexlab{}.
\newblock \showarticletitle{Hearing how smooth it looks: Selective attention and crossmodal perception in the arts}.
\newblock \bibinfo{journal}{\emph{Essays in Philosophy}} \bibinfo{volume}{13}, \bibinfo{number}{2} (\bibinfo{year}{2012}), \bibinfo{pages}{498--517}.
\newblock


\bibitem[Shneiderman(2004)]%
        {shneiderman2004designing}
\bibfield{author}{\bibinfo{person}{Ben Shneiderman}.} \bibinfo{year}{2004}\natexlab{}.
\newblock \showarticletitle{Designing for fun: how can we design user interfaces to be more fun?}
\newblock \bibinfo{journal}{\emph{interactions}} \bibinfo{volume}{11}, \bibinfo{number}{5} (\bibinfo{year}{2004}), \bibinfo{pages}{48--50}.
\newblock


\bibitem[Steen et~al\mbox{.}(2007)]%
        {steen2007early}
\bibfield{author}{\bibinfo{person}{Marc Steen}, \bibinfo{person}{Lottie Kuijt-Evers}, {and} \bibinfo{person}{Jente Klok}.} \bibinfo{year}{2007}\natexlab{}.
\newblock \showarticletitle{Early user involvement in research and design projects--A review of methods and practices}. In \bibinfo{booktitle}{\emph{23rd egos colloquium}}, Vol.~\bibinfo{volume}{5}. \bibinfo{pages}{1--21}.
\newblock


\bibitem[Still(2018)]%
        {still2018web}
\bibfield{author}{\bibinfo{person}{Jeremiah~D Still}.} \bibinfo{year}{2018}\natexlab{}.
\newblock \showarticletitle{Web page visual hierarchy: Examining Faraday's guidelines for entry points}.
\newblock \bibinfo{journal}{\emph{Computers in Human Behavior}}  \bibinfo{volume}{84} (\bibinfo{year}{2018}), \bibinfo{pages}{352--359}.
\newblock


\bibitem[Stokes et~al\mbox{.}(2025)]%
        {stokes2025write}
\bibfield{author}{\bibinfo{person}{Chase Stokes}, \bibinfo{person}{Kylie Lin}, {and} \bibinfo{person}{Cindy~Xiong Bearfield}.} \bibinfo{year}{2025}\natexlab{}.
\newblock \showarticletitle{Write, Rank, or Rate: Comparing Methods for Studying Visualization Affordances}.
\newblock \bibinfo{journal}{\emph{IEEE VIS}} (\bibinfo{year}{2025}).
\newblock


\bibitem[Team et~al\mbox{.}(2023)]%
        {team2023gemini}
\bibfield{author}{\bibinfo{person}{Gemini Team}, \bibinfo{person}{Rohan Anil}, \bibinfo{person}{Sebastian Borgeaud}, \bibinfo{person}{Jean-Baptiste Alayrac}, \bibinfo{person}{Jiahui Yu}, \bibinfo{person}{Radu Soricut}, \bibinfo{person}{Johan Schalkwyk}, \bibinfo{person}{Andrew~M Dai}, \bibinfo{person}{Anja Hauth}, \bibinfo{person}{Katie Millican}, {et~al\mbox{.}}} \bibinfo{year}{2023}\natexlab{}.
\newblock \showarticletitle{Gemini: a family of highly capable multimodal models}.
\newblock \bibinfo{journal}{\emph{arXiv preprint arXiv:2312.11805}} (\bibinfo{year}{2023}).
\newblock


\bibitem[Thakur et~al\mbox{.}(2025)]%
        {thakur2025judgingjudgesevaluatingalignment}
\bibfield{author}{\bibinfo{person}{Aman~Singh Thakur}, \bibinfo{person}{Kartik Choudhary}, \bibinfo{person}{Venkat~Srinik Ramayapally}, \bibinfo{person}{Sankaran Vaidyanathan}, {and} \bibinfo{person}{Dieuwke Hupkes}.} \bibinfo{year}{2025}\natexlab{}.
\newblock \bibinfo{title}{Judging the Judges: Evaluating Alignment and Vulnerabilities in LLMs-as-Judges}.
\newblock
\newblock
\showeprint[arxiv]{2406.12624}~[cs.CL]
\urldef\tempurl%
\url{https://arxiv.org/abs/2406.12624}
\showURL{%
\tempurl}


\bibitem[Wang(2024)]%
        {wang2024research}
\bibfield{author}{\bibinfo{person}{Di Wang}.} \bibinfo{year}{2024}\natexlab{}.
\newblock \showarticletitle{Research on Visual Hierarchy and Interactive Experience in Digital Media UI Design}. In \bibinfo{booktitle}{\emph{2024 5th International Conference on Intelligent Design (ICID)}}. IEEE, \bibinfo{pages}{133--136}.
\newblock


\bibitem[Wang(2025)]%
        {wang2025agenta}
\bibfield{author}{\bibinfo{person}{Dakuo et~al. Wang}.} \bibinfo{year}{2025}\natexlab{}.
\newblock \showarticletitle{AgentA/B: Automated and Scalable Web A/B Testing with Interactive LLM Agents}.
\newblock \bibinfo{journal}{\emph{arXiv preprint arXiv:2504.09723}} (\bibinfo{year}{2025}).
\newblock


\bibitem[Wang et~al\mbox{.}(2024)]%
        {wang2024aligned}
\bibfield{author}{\bibinfo{person}{Huichen~Will Wang}, \bibinfo{person}{Jane Hoffswell}, \bibinfo{person}{Victor~S Bursztyn}, \bibinfo{person}{Cindy~Xiong Bearfield}, {et~al\mbox{.}}} \bibinfo{year}{2024}\natexlab{}.
\newblock \showarticletitle{How aligned are human chart takeaways and llm predictions? a case study on bar charts with varying layouts}.
\newblock \bibinfo{journal}{\emph{IEEE Transactions on Visualization and Computer Graphics}} (\bibinfo{year}{2024}).
\newblock


\bibitem[Wang et~al\mbox{.}(2023)]%
        {wang2023large}
\bibfield{author}{\bibinfo{person}{Peiyi Wang}, \bibinfo{person}{Lei Li}, \bibinfo{person}{Liang Chen}, \bibinfo{person}{Zefan Cai}, \bibinfo{person}{Dawei Zhu}, \bibinfo{person}{Binghuai Lin}, \bibinfo{person}{Yunbo Cao}, \bibinfo{person}{Qi Liu}, \bibinfo{person}{Tianyu Liu}, {and} \bibinfo{person}{Zhifang Sui}.} \bibinfo{year}{2023}\natexlab{}.
\newblock \showarticletitle{Large language models are not fair evaluators}.
\newblock \bibinfo{journal}{\emph{arXiv preprint arXiv:2305.17926}} (\bibinfo{year}{2023}).
\newblock


\bibitem[Wang et~al\mbox{.}(2022)]%
        {wang2022self}
\bibfield{author}{\bibinfo{person}{Yizhong Wang}, \bibinfo{person}{Yeganeh Kordi}, \bibinfo{person}{Swaroop Mishra}, \bibinfo{person}{Alisa Liu}, \bibinfo{person}{Noah~A Smith}, \bibinfo{person}{Daniel Khashabi}, {and} \bibinfo{person}{Hannaneh Hajishirzi}.} \bibinfo{year}{2022}\natexlab{}.
\newblock \showarticletitle{Self-instruct: Aligning language models with self-generated instructions}.
\newblock \bibinfo{journal}{\emph{arXiv preprint arXiv:2212.10560}} (\bibinfo{year}{2022}).
\newblock


\bibitem[Wang et~al\mbox{.}(2019)]%
        {8809393}
\bibfield{author}{\bibinfo{person}{Yun Wang}, \bibinfo{person}{Adrien Segal}, \bibinfo{person}{Roberta Klatzky}, \bibinfo{person}{Daniel~F. Keefe}, \bibinfo{person}{Petra Isenberg}, \bibinfo{person}{Jörn Hurtienne}, \bibinfo{person}{Eva Hornecker}, \bibinfo{person}{Tim Dwyer}, {and} \bibinfo{person}{Stephen Barrass}.} \bibinfo{year}{2019}\natexlab{}.
\newblock \showarticletitle{An Emotional Response to the Value of Visualization}.
\newblock \bibinfo{journal}{\emph{IEEE Computer Graphics and Applications}} \bibinfo{volume}{39}, \bibinfo{number}{5} (\bibinfo{year}{2019}), \bibinfo{pages}{8--17}.
\newblock
\urldef\tempurl%
\url{https://doi.org/10.1109/MCG.2019.2923483}
\showDOI{\tempurl}


\bibitem[Wu et~al\mbox{.}(2024)]%
        {wu2024uiclip}
\bibfield{author}{\bibinfo{person}{Jason Wu}, \bibinfo{person}{Yi-Hao Peng}, \bibinfo{person}{Xin Yue~Amanda Li}, \bibinfo{person}{Amanda Swearngin}, \bibinfo{person}{Jeffrey~P Bigham}, {and} \bibinfo{person}{Jeffrey Nichols}.} \bibinfo{year}{2024}\natexlab{}.
\newblock \showarticletitle{UICLIP: a data-driven model for assessing user interface design}. In \bibinfo{booktitle}{\emph{Proceedings of the 37th Annual ACM Symposium on User Interface Software and Technology}}. \bibinfo{pages}{1--16}.
\newblock


\bibitem[Xiang et~al\mbox{.}(2024)]%
        {10.1145/3613904.3642481}
\bibfield{author}{\bibinfo{person}{Wei Xiang}, \bibinfo{person}{Hanfei Zhu}, \bibinfo{person}{Suqi Lou}, \bibinfo{person}{Xinli Chen}, \bibinfo{person}{Zhenghua Pan}, \bibinfo{person}{Yuping Jin}, \bibinfo{person}{Shi Chen}, {and} \bibinfo{person}{Lingyun Sun}.} \bibinfo{year}{2024}\natexlab{}.
\newblock \showarticletitle{SimUser: Generating Usability Feedback by Simulating Various Users Interacting with Mobile Applications}. In \bibinfo{booktitle}{\emph{Proceedings of the 2024 CHI Conference on Human Factors in Computing Systems}} (Honolulu, HI, USA) \emph{(\bibinfo{series}{CHI '24})}. \bibinfo{publisher}{Association for Computing Machinery}, \bibinfo{address}{New York, NY, USA}, Article \bibinfo{articleno}{9}, \bibinfo{numpages}{17}~pages.
\newblock
\showISBNx{9798400703300}
\urldef\tempurl%
\url{https://doi.org/10.1145/3613904.3642481}
\showDOI{\tempurl}


\bibitem[Zhang et~al\mbox{.}(2023)]%
        {zhang2023wider}
\bibfield{author}{\bibinfo{person}{Xinghua Zhang}, \bibinfo{person}{Bowen Yu}, \bibinfo{person}{Haiyang Yu}, \bibinfo{person}{Yangyu Lv}, \bibinfo{person}{Tingwen Liu}, \bibinfo{person}{Fei Huang}, \bibinfo{person}{Hongbo Xu}, {and} \bibinfo{person}{Yongbin Li}.} \bibinfo{year}{2023}\natexlab{}.
\newblock \showarticletitle{Wider and deeper llm networks are fairer llm evaluators}.
\newblock \bibinfo{journal}{\emph{arXiv preprint arXiv:2308.01862}} (\bibinfo{year}{2023}).
\newblock


\bibitem[Zheng et~al\mbox{.}(2023)]%
        {zheng2023judging}
\bibfield{author}{\bibinfo{person}{Lianmin Zheng}, \bibinfo{person}{Wei-Lin Chiang}, \bibinfo{person}{Ying Sheng}, \bibinfo{person}{Siyuan Zhuang}, \bibinfo{person}{Zhanghao Wu}, \bibinfo{person}{Yonghao Zhuang}, \bibinfo{person}{Zi Lin}, \bibinfo{person}{Zhuohan Li}, \bibinfo{person}{Dacheng Li}, \bibinfo{person}{Eric Xing}, {et~al\mbox{.}}} \bibinfo{year}{2023}\natexlab{}.
\newblock \showarticletitle{Judging llm-as-a-judge with mt-bench and chatbot arena}.
\newblock \bibinfo{journal}{\emph{Advances in Neural Information Processing Systems}}  \bibinfo{volume}{36} (\bibinfo{year}{2023}), \bibinfo{pages}{46595--46623}.
\newblock


\bibitem[Ziegler et~al\mbox{.}(2020)]%
        {ziegler2020finetuninglanguagemodelshuman}
\bibfield{author}{\bibinfo{person}{Daniel~M. Ziegler}, \bibinfo{person}{Nisan Stiennon}, \bibinfo{person}{Jeffrey Wu}, \bibinfo{person}{Tom~B. Brown}, \bibinfo{person}{Alec Radford}, \bibinfo{person}{Dario Amodei}, \bibinfo{person}{Paul Christiano}, {and} \bibinfo{person}{Geoffrey Irving}.} \bibinfo{year}{2020}\natexlab{}.
\newblock \bibinfo{title}{Fine-Tuning Language Models from Human Preferences}.
\newblock
\newblock
\showeprint[arxiv]{1909.08593}~[cs.CL]
\urldef\tempurl%
\url{https://arxiv.org/abs/1909.08593}
\showURL{%
\tempurl}


\bibitem[Zieglmeier and Lehene(2021)]%
        {zieglmeier2021designing}
\bibfield{author}{\bibinfo{person}{Valentin Zieglmeier} {and} \bibinfo{person}{Antonia~Maria Lehene}.} \bibinfo{year}{2021}\natexlab{}.
\newblock \showarticletitle{Designing trustworthy user interfaces}. In \bibinfo{booktitle}{\emph{Proceedings of the 33rd Australian Conference on Human-Computer Interaction}}. \bibinfo{pages}{182--189}.
\newblock


\end{thebibliography}

\newpage
\appendix

\section{Opportunities (Extended)}
Building on our benchmark, we are extremely excited by the possible opportunities to expand on the capabilities of MLLM as a judge for interface testing. Most notably, this benchmark can be used to create a fine-tuned model or in alignment training. Reinforcement learning from human feedback (RLHF) is a specialized fine-tuning technique that would most align with this benchmark. Whether using this benchmark in isolation or in tandem with other similar datasets, there is an opportunity to create a fine-tuned model that will better supplement user research.

\begin{figure}[h]
\centering
\includegraphics[width=1.0\linewidth]{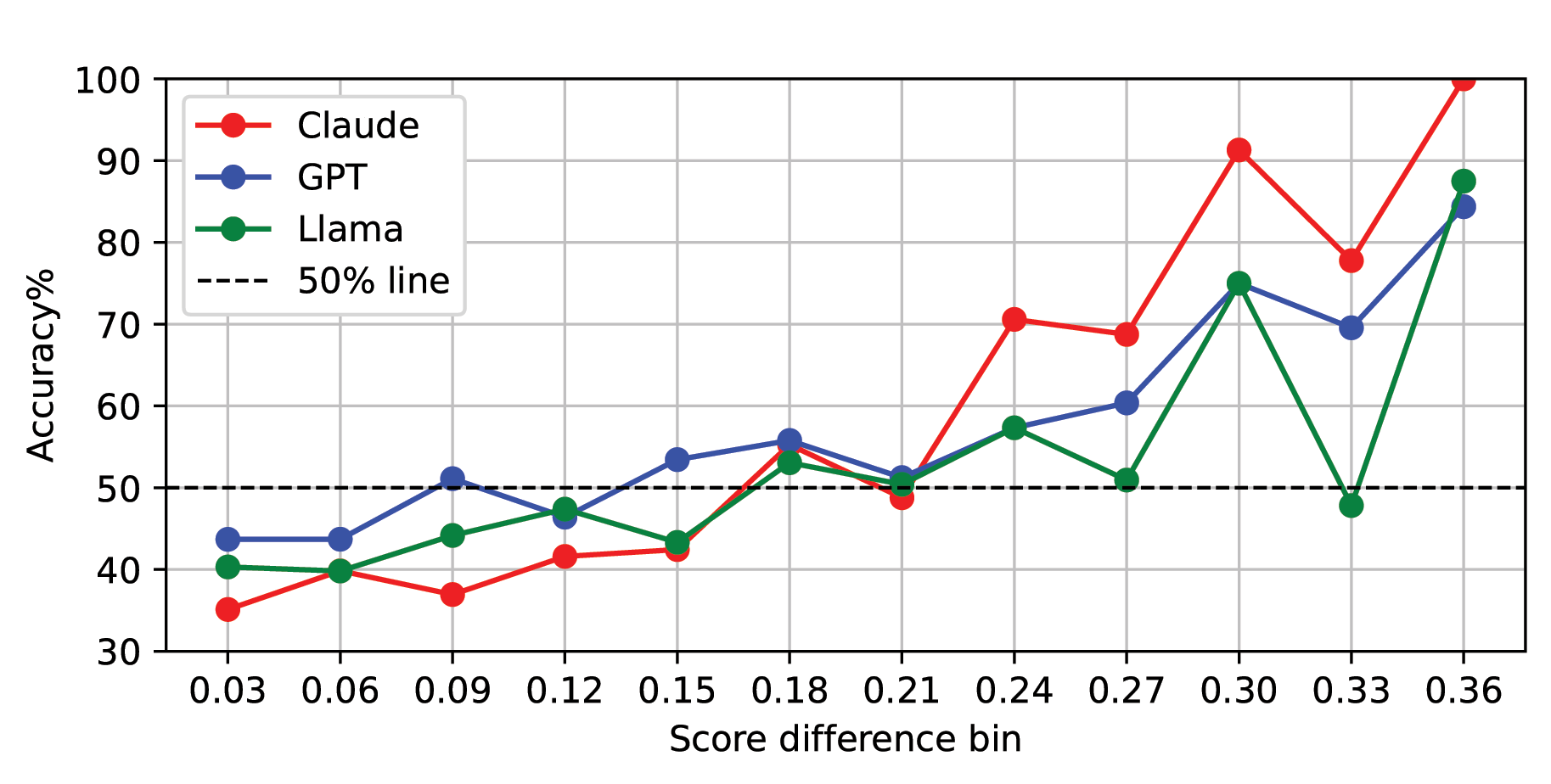}
\caption{%
Results when pairwise data is created for MLLMs and humans in the same way. For both human and MLLM evaluation sets, we take the absolute value gathered in task 1, compare the two absolute values, and take the higher of the two. Then we compare the human results to the MLLM results to get the agreement score presented here.
}

\label{fig:PairwiseAgreement2}
\end{figure} 

Secondly, there is also the opportunity to expand on the benchmark by adding more screens and/or more tasks. Increasing the amount and diversity of screens can uncover more trends that were not uncovered in this study. For example, this study focused on mobile and desktop first UIs, but UIs exist for tablets, televisions, entertainment consoles, and many other devices. Expanding the benchmark to include a diverse set of UIs will create a more holistic benchmark. Furthermore, user research is not limited to just Likert-scale scores and A/B testing. The benchmark can be expanded by adding usability studies, surveys, need-finding, etc. Expanding the data in such a way will further test MLLM's efficacy in UI tasks and create a more expansive benchmark for fine-tuned models.

\begin{figure*}[h!]
\centering
\includegraphics[width=0.6\linewidth]{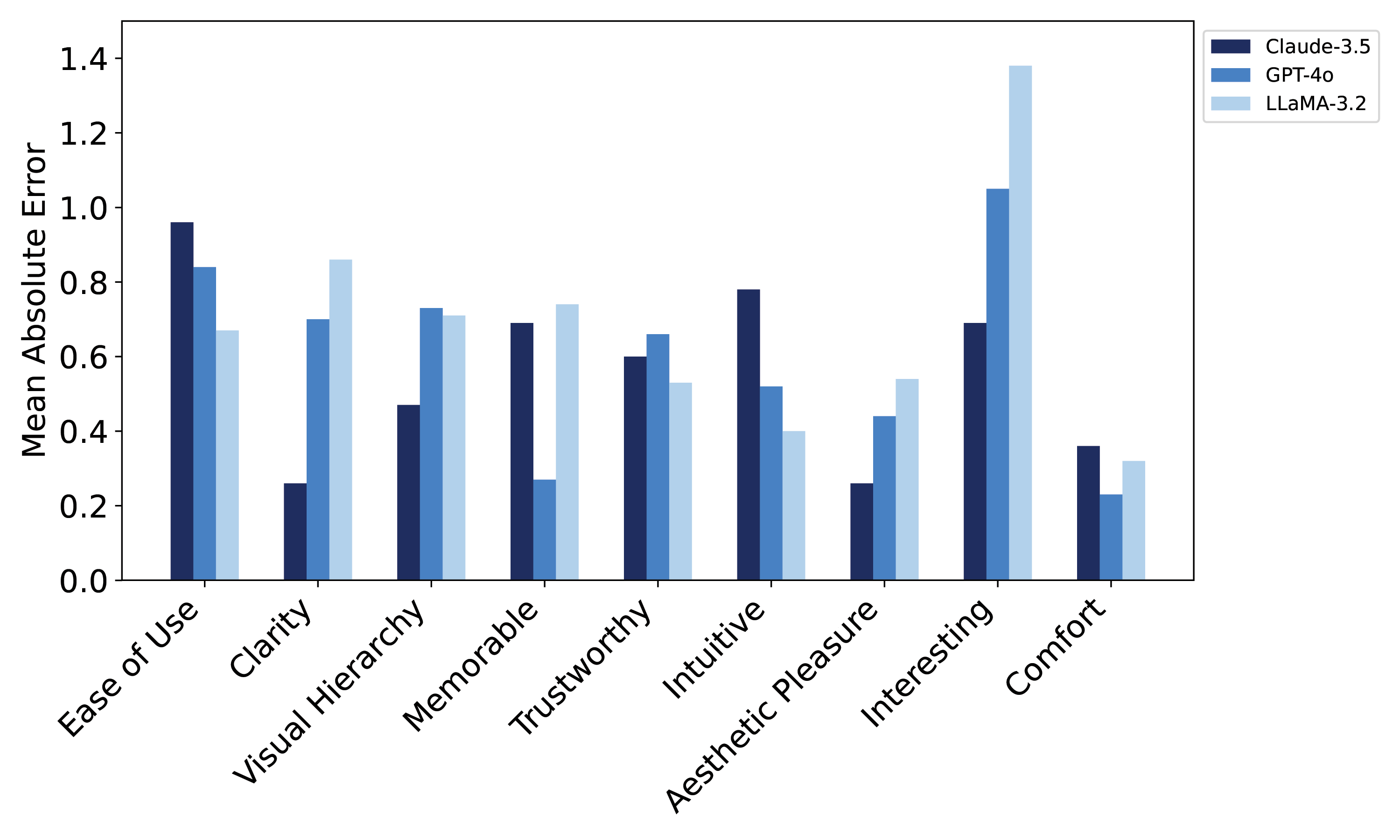}
\caption{%
\textbf{Mean Absolute Error} (lower is better): This result shows that the mean absolute error for all models consistently is less than 1, with "interesting" being the exception. These results show that the models can reliably score within one point of the human scores when given a Likert scale UI question. }
\label{fig:MAEPLot}
\end{figure*}

There is also the opportunity to study the efficacy of more models and prompting methods. This study revolved mostly around GPT-4o, Claude 3.5 Sonnet, and Llama 3.2; running the same tasks on other models like Google's Gemini, Qwen, or even GPT-4o mini could unveil even further outcomes and trends. Future studies can utilize the framework established in this study to test a wider range of models. Similarly, there is an opportunity to explore the role that prompting strategies can play in MLLM as a UI judge. For this reason, future studies can explore chain-of-thought and few shot learning methods to explore the impact that these methods have on outputs.

Finally, considering that UI design is an iterative process, there is an opportunity for future studies to compare UI designs to their earlier iterations. In the pairwise portion of the study, different UI designs were compared with one another. Instead, there is an opportunity to explore the impact of changing a single color, text, or even the layout, and comparing it with an earlier iteration to measure the degree that small changes within the same design change human and MLLM judgments.

\begin{figure*}[ht]
\vspace{5mm}
    \centering
    \hfill
    \hspace{-3mm}
    \subfigure[Browsing \& Discovery]{
        \includegraphics[width=0.37\textwidth]{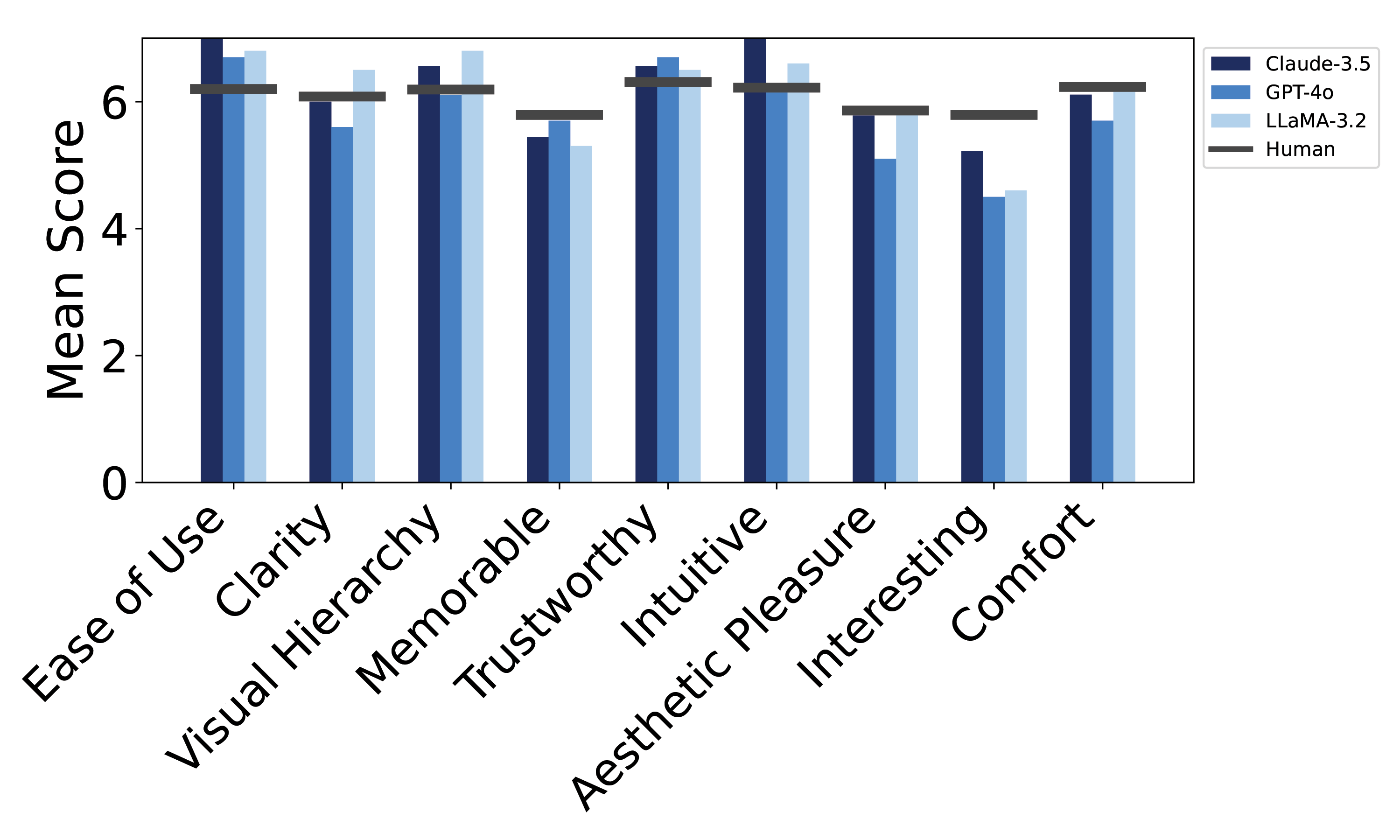}
    }
    \hspace{-12.8mm}
    \subfigure[Confirmation \& Feedback]{
        \includegraphics[width=0.37\textwidth]{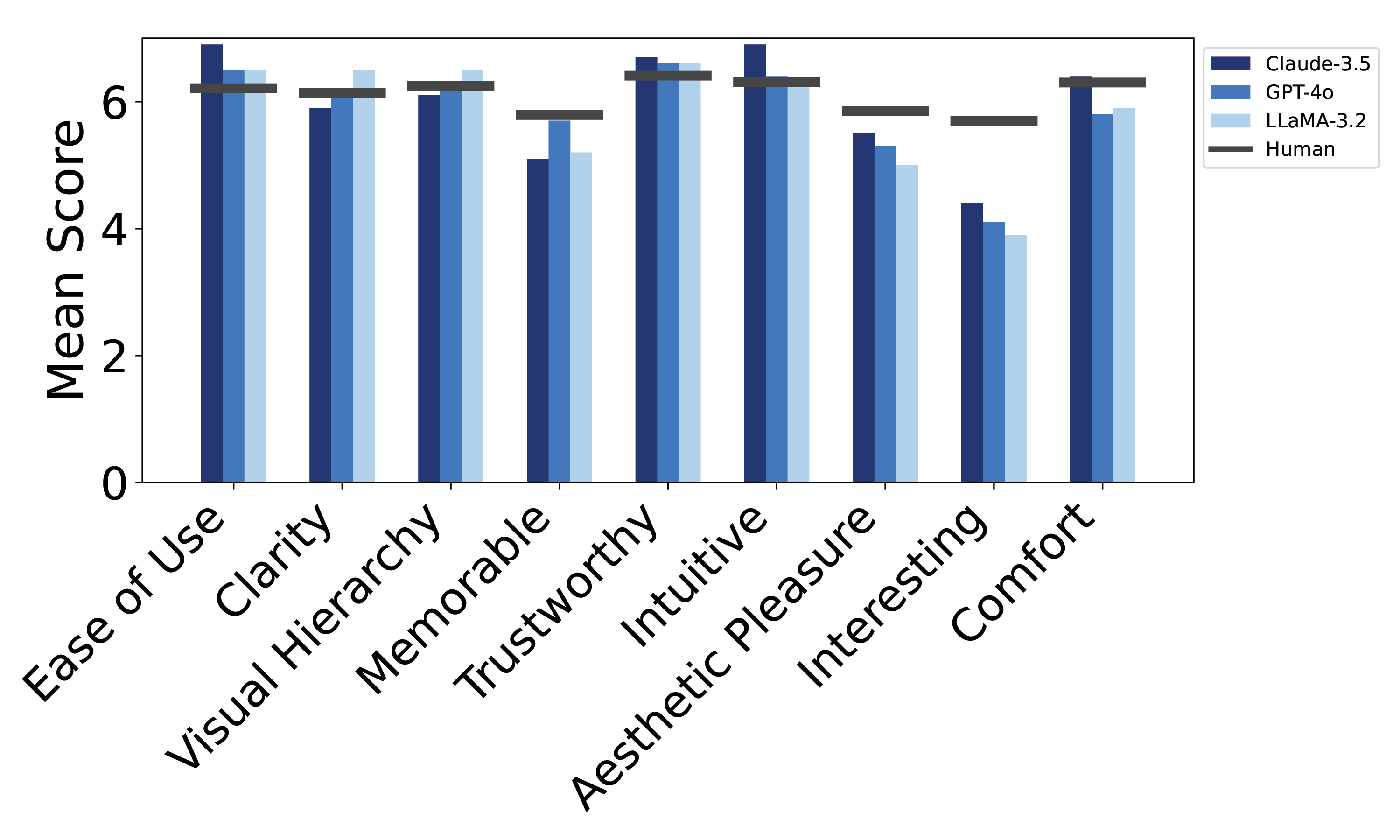}
    }
    \hspace{-12.8mm}
    \subfigure[Communication \& Engagement]{
        \includegraphics[width=0.37\textwidth]{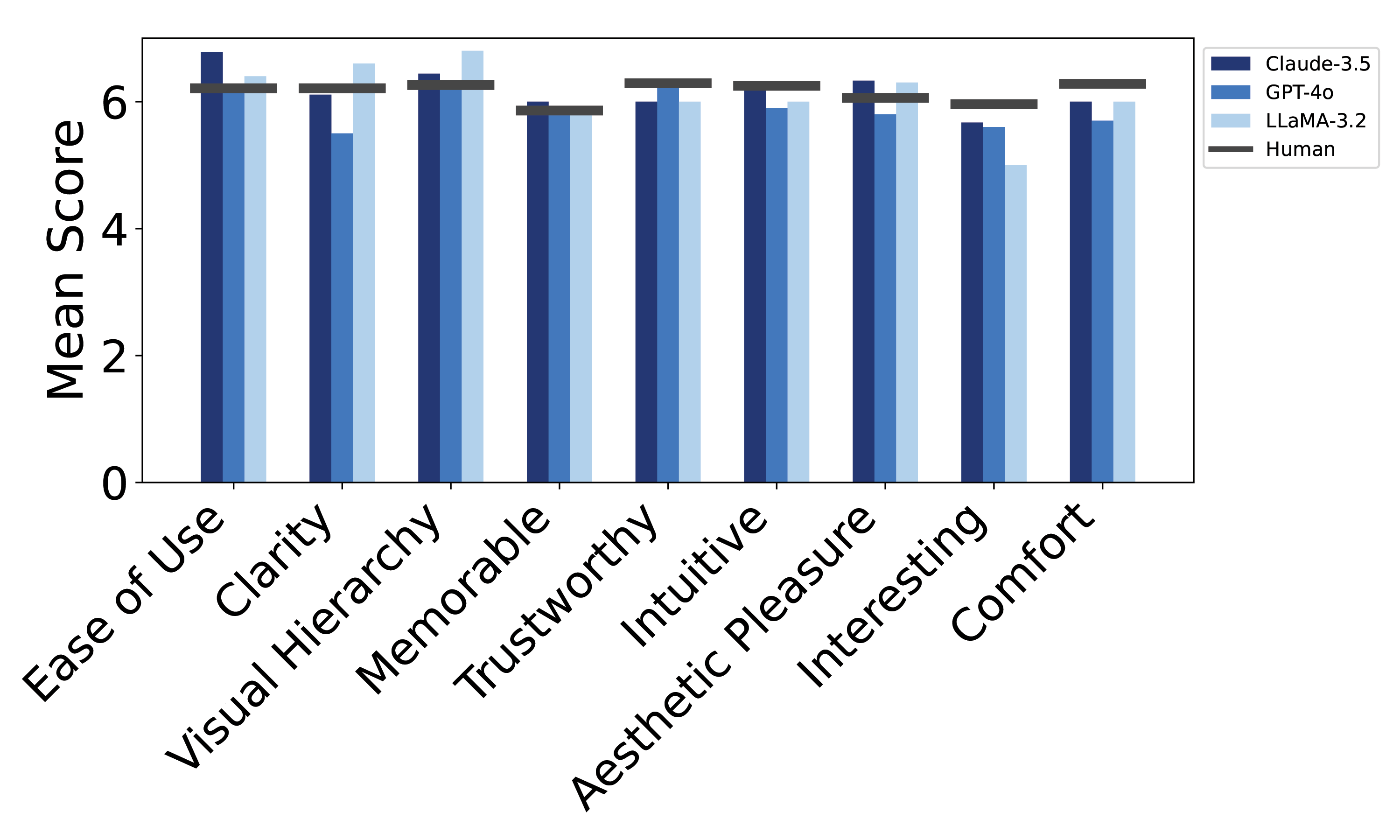}
    }

    \caption{%
    Three grouped bar charts comparing mean factor scores from humans vs. each model. Human score marked with black line, model scores represented by different bars. Trends show that MLLMs are able to effectively approximate human scores. Also shows how model scoring distributions align with human scores across Browsing \& Discovery, Confirmation \& Feedback, and Communication \& Engagement. }
    \label{fig:LikertBars}
\end{figure*}

\section{Additional Results \& Prompts}
The additional results shown reflect further visualizations of data presented in the main body of the paper. \cref{fig:MAEPLot} and \cref{fig:LikertBars} represent how well MLLMs did on Task 1, with the former representing the mean absolute error (MAE) and the latter being a visual summary of \cref{table:LikertTable}. Furthermore works like \cref{fig:PairwiseAgreement2}, represent potential ablation study data. 
Here we also present the prompts used to have the MLLMs evaluate the UIs in the same way that the humans did. \cref{fig:prompt-UI1} asks the MLLMs to evaluate the UIs on a 7-point Likert scale on the UI factors presented in \cref{table:Factors}. \cref{fig:prompt-pairwise} prompts the MLLMs to compare two UIs to one another on the aforementioned factors and choos a "winner."

\begin{figure*}[ht]
\centering
\begin{promptbox}{UI Evaluation Task Pt.1}\vspace{2mm}
\begin{small}
\texttt{\texttt{"You are an average user brought in to do human testing. For the given UI image, 
evaluate the following nine qualities. 
For each, give one of the following ratings: 1 (strongly disagree), 2 (disagree), 3 (slightly disagree), 4 (neutral), 5 (slightly agree), 6 (agree), or 7 (strongly agree), followed by a short rationale. **Format your response exactly like this:**
\begin{description}
    \item[] "1. **The UI is easily remembered**: **[x]/7** Your rationale here."
    \item[] "2. **The UI appears trustworthy**: **[x]/7** Your rationale here."
    \item[] "3. **The UI is aesthetically pleasing**: **[x]/7** Your rationale here."
    \item[] "..."
    \item[] "Do NOT use any other formatting like parentheses or dashes. Only use the format shown."
    \item[] Evaluate the following:
   \item[]"The UI is easily remembered.",
   \item[]  "The UI appears trustworthy.",
   \item[]  "The UI is aesthetically pleasing.",
   \item[]  "The UI is intuitive.",
   \item[]  "The UI is interesting.",
   \item[]  "I feel comfortable with the UI.",
   \item[]  "The UI looks easy to use.",
   \item[]  "The layout is uncluttered",
   \item[]  "The UI has a clear visual hierarchy"
\end{description}
}}
\begin{center}
\texttt{[UI Image]}
\end{center}
\end{small}
\vspace{2mm}
\end{promptbox}
\vspace{-2mm}
\caption{Instructions provided to the MLLMs to evaluate the UI in \cref{sec:exp-score-pred}}
\label{fig:prompt-UI1}
\end{figure*}

\begin{figure*}[ht]
\centering
\begin{promptbox}{UI Pairwise Evaluation Prompt}
\vspace{2mm}
\begin{small}
You are an average user evaluating a user interface.\\

You will be shown two UI screenshots:\\

UI-A:
\begin{center}
\texttt{[UI Image A]}
\end{center}

UI-B:
\begin{center}
\texttt{[UI Image B]}
\end{center}

Your task is to determine which UI an ordinary person would prefer for 10 evaluation criteria.\\

For each criterion, write:\\
    <\textcolor{googlegreen}{criterion}>[Criterion Text]</\textcolor{googlegreen}{criterion}>\\
    <\textcolor{googleblue}{result}>[UI-A or UI-B]</\textcolor{googleblue}{result}>\\
    <\textcolor{googlered}{reason}>[your reasoning in less than 50 words]</\textcolor{googlered}{reason}>\\

Here are the evaluation criterion:\\
    \quad* The UI is easily remembered\\
    \quad* The UI appears trustworthy\\
    \quad* The UI is aesthetically pleasing\\
    * ... \\
\end{small}
\vspace{2mm}
\end{promptbox}
\vspace{-2mm}
\caption{Instructions provided to the MLLMs for pairwise comparison in ~\cref{sec:exp-pairwise-comparison}.}
\label{fig:prompt-pairwise}
\end{figure*}

\end{document}